\documentclass[twocolumn,aps,superscriptaddress,pre]{revtex4}

\usepackage{amsmath,amssymb,graphicx}
\usepackage{algorithmic}
\usepackage{enumerate}
\usepackage{color}
\usepackage{soul,xcolor}  %added for strikethrough option 
\setstcolor{red} %strikethrough color, use \st{}

\DeclareMathOperator{\sech}{sech}

\graphicspath{{./fig-jpg/}{./fig-ps/}}

\begin{document}

\title{Reduced dynamics for one and two dark soliton stripes\\
in the defocusing nonlinear Schr\"odinger equation: a variational approach}

\author{L. A. Cisneros-Ake}
\affiliation{Departamento de Matem\'aticas, ESFM, Instituto Polit\'{e}cnico Nacional,
Unidad Profesional Adolfo L\'{o}pez Mateos Edificio 9, 07738 Cd.~de
M\'exico, M\'{e}xico}
%\email{cisneros@esfm.ipn.mx}

\author{R. Carretero-Gonz{\'a}lez}
\affiliation{Nonlinear Dynamical Systems Group,\footnote{\texttt{URL}: http://nlds.sdsu.edu}
Computational Sciences Research Center, and
Department of Mathematics and Statistics,
San Diego State University, San Diego, California 92182-7720, USA}

\author{P. G. Kevrekidis}
%\email{kevrekid@math.umass.edu}
\affiliation{Department of Mathematics and Statistics, University of Massachusetts,
Amherst, Massachusetts 01003-4515 USA}

\begin{abstract}

We study the dynamics and pairwise interactions of dark soliton
stripes in the two-dimensional defocusing nonlinear Schr\"odinger 
equation.
By employing a variational approach we reduce the dynamics for dark 
soliton stripes to a set of coupled one-dimensional ``filament'' 
equations of motion for the position and velocity of the stripe.
The method yields good qualitative agreement with the numerical
results as regards the transverse instability of the stripes. We
propose a phenomenological amendment that also significantly
improves the quantitative agreement of the method with the computations.
Subsequently, the method is extended for a pair of symmetric dark soliton
stripes that include the mutual interactions between the filaments.
The reduced equations of motion are compared with a recently proposed
adiabatic invariant method and its corresponding findings and are found
to provide a more adequate representation of 
the original full dynamics for a wide range of cases encompassing
perturbations with long and short wavelengths, and combinations
thereof.

\end{abstract}

\maketitle

%%%%%%%%%%%%%%%%%%%%%%%%%%%%%%%%%%%%%%%%%%%%%%%%%%%%%%%%%%%%%%%%%%%%%%%%%%%%%
\section{Introduction}
\label{sec:intro}
%%%%%%%%%%%%%%%%%%%%%%%%%%%%%%%%%%%%%%%%%%%%%%%%%%%%%%%%%%%%%%%%%%%%%%%%%%%%%

In the past two decades, the study of coherent structures in the
form of dark solitons has been a theme pervading a wide range of
areas within Physics. Early examples were more focused
on classical physics (including mechanical~\cite{mech} and electrical~\cite{el} lattices), 
nonlinear optical~\cite{Kivshar-LutherDavies}, as well
as magnetic systems~\cite{mf}. More recently, however, there has been
a host of additional systems including notably a wide variety
of experiments in atomic Bose-Einstein condensates summarized, e.g.,
in Refs.~\cite{djf,SIAMbook}, but also realizations in electromagnetic~\cite{gv},
hydrodynamic~\cite{ww}, acoustic~\cite{chong}, plasma~\cite{plasma}
and exciton-polariton~\cite{polar} systems among others.

A major thrust within the subject has been offered by the extensive
experimental accessibility to such nonlinear waves offered in the
context of atomic (but also exciton-polariton) condensates. Here,
there has been a variety of techniques of formation of the structures,
including wave interference~\cite{ourmarkus2,ourmarkus3}, phase
imprinting~\cite{becker} and rapid dragging of laser beams through
a trapped BEC~\cite{engels}. Additionally, the structures have
offered a variety of intriguing insights in the dynamics through
their potential dynamical instabilities (and their avoidance
through suitable manipulation of length scales~\cite{infrared})
which may lead to a variety of vortical (in 2D) and vortex line/ring
structures (in 3D), as has also been documented
experimentally~\cite{watching,jeff,becker2,lamporesi}.
More recently extensions to multi-component settings have been
pursued in the form of dark-bright and dark-dark solitonic
states~\cite{revip} and even spinorial realizations of such structures
have recently been identified~\cite{bersano}. At the same time,
there has been a growing interest in applications including
possibilities of atomic matter-wave interferometers~\cite{in1} and
proposals for the use of dark solitons as qubits in BECs~\cite{qub}.

The study of transverse (``snaking'') dynamical instabilities of
dark solitons in higher dimensional settings has been a topic of
extensive interest since early on~\cite{kuzne}, with much of
the early work summarized in the review~\cite{kidep}. Recently,
there has been a surge of further interest in the
subject~\cite{aipaper,aipaper2,aipaper3} fueled by an adiabatic
invariant (AI) based insight enabling the derivation of effective
equations for the dark soliton stripes (in 2D) and planes (in 3D),
but also the ability of this methodology to tackle ring
(in 2D, but also in 3D in the form of spherical) solitons~\cite{Kivshar-Yang}.
This methodology was not only seen to have the right long wavelength limit
(a fundamental pre-requisite for such a theory). Additionally
in many cases, including those of ring and spherical solitons, it provided
with unprecedented accuracy and simplicity analytical approximations
for the frequencies of vibration and destabilization of the coherent
structures that were tested in both linearization (spectral) computations,
as well as in the fully nonlinear dynamics of the system.

Nevertheless, there is reason to believe that the theory can be
{\it further} improved. On the one hand, while the above AI
methodology captures the correct long wavelength limit, it does not
a priori capture the restabilization of perturbations of large wavenumbers
(above a certain $k_{c}$). Moreover, as it stands, the theory is developed
solely for the evolutionary dynamics of the center of the dark solitonic
stripes (or planes), but does not arise as a coupled theory for the
evolution of the center and the width (or velocity) of the structure.
For these reasons, but also in order to obtain a theory with a definitive
Lagrangian framework (avoiding the issue of identification of
canonical variables and) enabling the systematic
derivation of Euler-Lagrange equations of motion, we develop herein
an alternative variational approach.

At the heart of our analysis lies a two-dimensional generalization of the classic 
work of Ref.~\cite{Kivshar-Krolikowski} considering the variational characterization
of the dynamics of a dark soliton in 1D. Here, we endow the dark soliton
stripe (DSS) with a center, a width and a speed that are transversely
dependent (as in the recent AI work of Refs.~\cite{aipaper,aipaper2,aipaper3}),
yet we substitute the relevant ansatz of one or two stripes
in the Lagrangian of the model. The subsequent derivation of the Euler-Lagrange
equations of motion for two dynamical variables (e.g. position and
velocity) constitutes the basis for our further analysis of the theory
and its improved, as we will show, agreement with the full numerical
results. We discuss some of the limitations of the method, such
as its qualitative but not quantitative capture of the restabilization
of large wavenumbers and offer relevant amendments that provide
the optimal characterization, to our knowledge, of the motion of
single and multiple dark solitonic stripes available to date. 
The manuscript is structured as follows.
Section~\ref{sec:1DSS} describes the VA methodology and puts forwards
the reduced equations of motion for a single DSS and for two 
(symmetrically displaced) interacting DSSs.
Section~\ref{sec:numerics} is devoted to testing the validity of the
VA approach and to compare its predictions against a previous filament
technique based on adiabatic invariance (AI)~\cite{aipaper,aipaper2,aipaper3}.
Finally, in Sec.~\ref{sec:conclusions} we summarize our results and
give possible avenues for future research.

%%%%%%%%%%%%%%%%%%%%%%%%%%%%%%%%%%%%%%%%%%%%%%%%%%%%%%%%%%%%%%%%%%%%%%%%%%%%%
\section{Variational Approach for One Stripe}
\label{sec:1DSS}
%%%%%%%%%%%%%%%%%%%%%%%%%%%%%%%%%%%%%%%%%%%%%%%%%%%%%%%%%%%%%%%%%%%%%%%%%%%%%

Our starting point will be the prototypical 2D NLS
model~\cite{Kivshar-LutherDavies,SIAMbook}:
\begin{equation}
\mathrm{i} u_t+\frac{1}{2}u_{xx}-\frac{\alpha}{2}u_{yy}-\left(|u|^2-u_0^2\right)u=0, \label{eq:NLS_2D}
\end{equation}
where $u_0$ is the magnitude of the background and $\alpha=\mp 1$ accounts for 
the elliptic and hyperbolic NLS cases. 
In our numerical results we will focus on the elliptic NLS case (i.e., $\alpha=-1$).
Nonetheless, for genericity's sake, we keep $\alpha$ in our analysis herein.
In the 1D case ($\alpha=0$), the NLS admits the following exact traveling dark soliton
solution:
\begin{equation}
u_{\rm 1D}\left(x,t\right)= \left[B\tanh \left(B\left(x-vt\right)\right)+\mathrm{i} A\right]e^{\mathrm{i} kx},
\label{eq3}
\end{equation}
where $A^2+B^2=u_0^2-\frac{k^2}{2}$ and $A=v-k$. We recall that $k$ accounts 
for the velocity of the background while $v$ denotes the velocity of the dark
soliton itself.
To variationally follow the dark soliton as a quasi-1D stripe (or filament) in the 2D 
NLS~(\ref{eq:NLS_2D}), we consider the 2D extension $u_{\rm 2D}=u_{\rm 2D}(x,y,t)$ by
allowing the position $X(y,t)$ of the dark soliton to be a function of $y$ and $t$ 
(in the spirit also of earlier works such as Refs.~\cite{aipaper,aipaper2,aipaper3}),
while keeping a stationary background ($k=0$), as follows:
\begin{equation}
\label{eq4}
u_{\rm 2D}= B\left(y,t\right)\tanh \left(D\left(y,t\right)\left(x-X\left(y,t\right)\right)\right)+\mathrm{i} A\left(y,t\right),
\end{equation}
where we consider the inverse width of the DSS $D$ to be, in general, 
decoupled (in the dark soliton core) from the background level $B$.
Nonetheless, we enforce $A^2+B^2=u_0^2=$constant to keep the DSS from
affecting the tails at $x=\pm\infty$.
Note that when $B=D$ and $v=X_{t}=A$, the DSS ansatz~(\ref{eq4}) reduces to 
the exact dark soliton (\ref{eq3}) mounted on a stationary background ($k=0$).

The NLS~(\ref{eq:NLS_2D}) is derived from the renormalized (in the $x$-direction) Lagrangian:
\begin{eqnarray}
L_{\rm 2D}=\int_{-\infty}^{\infty} L_y\, dy, 
\label{eq:L2D}
\end{eqnarray}
where the averaged (i.e., integrated) Lagrangian along the $x$-direction may be written as
\begin{eqnarray}
L_y &=& \int_{-\infty}^{\infty} \left[\frac{\mathrm{i}}{2}\left(u^*u_t-uu_t^*\right)\left(1-\frac{u_0^2}{|u|^2}\right)
\right.
\notag
\\
\label{eq5a}
&&
\left.
-\frac{1}{2}|u_x|^2+\frac{\alpha}{2}|u_y|^2-\frac{1}{2}\left(|u|^2-u_0^2\right)^2\right]dx. 
\end{eqnarray}
We note that the term $\left(1-{u_0^2}/{|u|^2}\right)$ is introduced to 
renormalize the momentum while the term $|u|^2-u_0^2$ renormalizes the power.
This renormalization in introduced so that the (1D) averaged Lagrangian $L_y$ 
converges when evaluated over the 1D dark 
soliton~(\ref{eq3})~\cite{Kivshar-Yang,Kivshar-Krolikowski}.

Let us now evaluate the 2D Lagrangian~(\ref{eq:L2D}) over the ansatz~(\ref{eq4})
by assuming that the dark soliton moves locally and thus it does not affect the
tails (background) in $A$ and $B$. Namely, we will use the following approximation
for the $y$-derivative of the 2D ansatz~(\ref{eq4}):
\begin{eqnarray}
u_y&\approx&B\left[D(x-X)\right]_y \sech^2 \left(D(x-X)\right)
\notag
\\
\label{eq5b}
&\approx&
B\left[D_y(x-X)-DX_{y}\right]\sech^2 \left(D(x-X)\right). 
~~
\end{eqnarray}
Alternatively, the above approximation can be thought of assuming a generic 
traveling profile $u\approx f\left[D(t,y)(x-X(t,y))\right]$ that yields,
using the chain rule, $u_y=u_x\left[\frac{D_y}{D}(x-X)-X_{y}\right]$. 
Finally, after evaluating the expression~(\ref{eq5b}) we enforce that, 
due to the balance outside the dark soliton core region, 
$D=B$ and $A^2+B^2=u_0^2$ which leads to the simplified average Lagrangian: 
\begin{eqnarray}
L_y &=& 2X_{t}\left(-AB+u_0^2 \arctan{\frac{B}{A}}\right)-\frac{4}{3}B^3
\notag
\\
&&+\alpha\frac{\pi^2-6}{18B}B_y^2+\frac{2\alpha}{3}B^3X_{y}^2.  \label{eq6}
\end{eqnarray}
As per the Euler-Lagrange prescription, we now take variations in the 
variables $X$ and $B$ which yield, respectively:
\begin{eqnarray}
%\delta X: 
B_t&=&-\alpha\frac{A}{3B^2}\left(B^3X_{y}\right)_y, \label{eq7}
\end{eqnarray}
and

\begin{eqnarray}
%\delta B: 
\frac{\alpha}{9}(\pi^2-6)\left(\frac{B_y}{B}\right)_y
&=&\frac{4B^2}{A}X_{t}
-\alpha\frac{\pi^2-6}{18B^2}B_y^2
\notag
\\
&&
\label{eq8}
-4B^2
+2\alpha B^2X_{y}^2. 
~~
\end{eqnarray}
These coupled partial differential equations (PDEs) represent the
equations of motion for the 2D ansatz (\ref{eq4}) in terms of the
variables $A$, $B$, and $X$. Using the relation $A^2+B^2=u_0^2$,
we can decouple and rewrite these coupled PDEs in terms of
the variables $A$ and $X$:
\begin{eqnarray}
A_t&=&\alpha\frac{B^2}{3}X_{yy}-\alpha AA_yX_{y}, 
\label{eq9}
\\[2.0ex]
X_{t}&=&A-\alpha \frac{\pi^2-6}{36B^4}A^2A_{yy}-\frac{\alpha}{2}AX_{y}^2
\notag
\\
&&
\label{eq10}
-\alpha\frac{\pi^2-6}{36B^6}AA_y^2\left(u_0^2+\frac{A^2}{2}\right), 
\end{eqnarray}
where $B^2=u_0^2-A^2$.

%%%%%%%%%%%%%%%%%%%%%%%%%%%%%%%%%%%%%%%%%%%%%%%%%%%%%%%%%%%%%%%%%%%%%%%
\begin{figure}[tbp]
\begin{center}
\includegraphics[width=0.9\columnwidth]{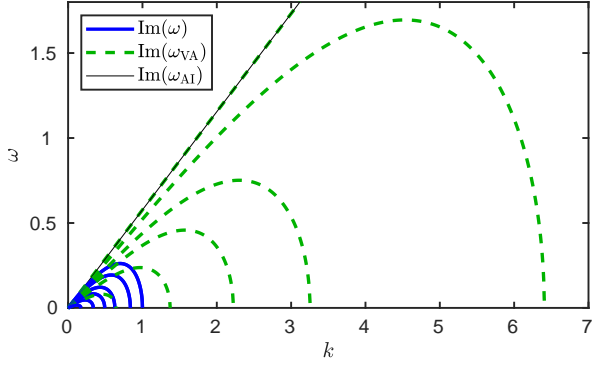}
\caption{(Color online) 
Stability spectrum for the DSS. Depicted is the eigenfrequency
$\omega$ as a function of the transverse wavenumber $k$
for $\mu=1$. The solid blue line depicts the unstable modes
from full NLS numerics corresponding to the imaginary part of
$\omega$.
The VA prediction of Eq.~(\ref{eq13}) is depicted by the green dashed curves
while the AI prediction (\ref{eq:dispAI}) is depicted by the thin
black line.
The different curves represent different stripe velocities $\eta$
corresponding, from top to bottom, to $\eta=0$, 0.4, 0.6, 0.7, 0.8, and 0.9.
%velspan2=[0,0.4,0.6,0.7,0.8,0.9];
%
}
\label{fig:spectrum_orig}
\end{center}
\end{figure}
%%%%%%%%%%%%%%%%%%%%%%%%%%%%%%%%%%%%%%%%%%%%%%%%%%%%%%%%%%%%%%%%%%%%%%%

Let us now study the (linear) stability for the reduced VA PDE
system of Eqs.~(\ref{eq9}) and (\ref{eq10}).
To this end, we linearize the system around the equilibrium configuration
$A=\eta$ and $X=0$ by considering $A= \eta+a$ 
and $X=\eta t+\xi$ for $a,\xi \ll 1$ to find:
\begin{eqnarray}
a_t &=& \alpha\frac{u_0^2-\eta^2}{3}\xi_{yy}, 
\label{eq11} 
\\
\label{eq12}
\xi_{t} &=& a-\frac{\alpha}{36}(\pi^2-6)\frac{\eta^2}{(u_0^2-\eta^2)^2}a_{yy}. 
\end{eqnarray}
Linear stability of plane waves $\left[a(y,t),\xi(y,t)\right]=\left[a_0 e^{\mathrm{i} (ky-\omega t)},b_0 e^{\mathrm{i} (ky-\omega t)}\right]$ the dispersion relation:
\begin{eqnarray}
\label{eq13}
\omega^2=\frac{\eta^2-u_0^2}{3}k^2\left[1+\frac{\alpha}{36}(\pi^2-6)\frac{\eta^2}{(u_0^2-\eta^2)^2}k^2\right]. 
\end{eqnarray}
As depicted in Fig.~\ref{fig:spectrum_orig} (see green dashed curves), Eq.~(\ref{eq13}) 
describes a dispersion relation $\omega=\omega(k)$ that has qualitatively the
correct shape when compared to the stability of the full (original) NLS 
model (\ref{eq:NLS_2D}) (see solid curves), obtained numerically.
However, we note that, despite  having the correct trend, the dispersion 
curves fail to accurately capture quantitatively the
actual growth rates and wavenumber cutoffs for stability (i.e., 
wavenumbers $k_{c}$ such that $\omega(k_{c})=0$).
This qualitative shape improves on the previous reduced description for the
dynamics of DSSs based on the AI
assumption~\cite{aipaper,aipaper2,aipaper3}. The reduced AI PDE for DSSs
yields, for $\eta=0$, the following (linear) dispersion relation~\cite{aipaper3}:
\begin{eqnarray}
\label{eq:dispAI}
\omega^2=-\frac{u_0^2}{3}k^2,
\end{eqnarray}
which simply predicts a linear growth (see thin black line in 
Fig.~\ref{fig:spectrum_orig}) of the instability rates as the wavenumber
increases.
Therefore, inspired by the correct qualitative shape of the dispersion curves
from the VA reduced model, we now propose a modification of the VA model so 
as to also quantitatively match more adequately the dispersion curves.

%%%%%%%%%%%%%%%%%%%%%%%%%%%%%%%%%%%%%%%%%%%%%%%%%%%%%%%%%%%%%%%%%%%%%%%
\begin{figure}[tbp]
\begin{center}
\includegraphics[width=0.9\columnwidth]{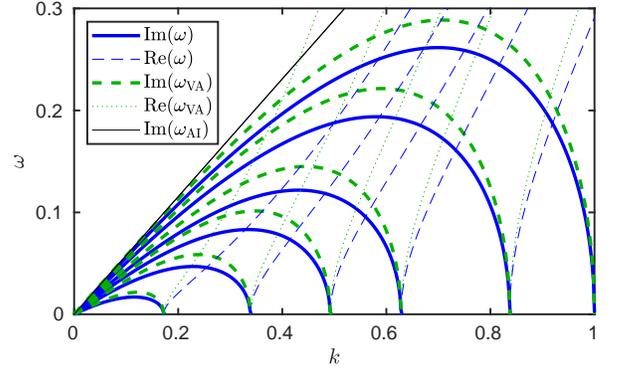}
\caption{(Color online) 
Stability spectrum for the DSS. Depicted is the eigenfrequency
$\omega$ as a function of the transverse wavenumber $k$
for $\mu=1$. 
Same notation and layout as in Fig.~\ref{fig:spectrum_orig}
where now the thick green dashed line corresponds to the 
{\em improved} VA prediction of Eq.~(\ref{eq:dispVAimproved}).
For completeness we also show the real part of the eigenfrequency
for the full NLS system (thin blue dashed curves) and the
improved VA (green dotted line).
}
\label{fig:spectrum}
\end{center}
\end{figure}
%%%%%%%%%%%%%%%%%%%%%%%%%%%%%%%%%%%%%%%%%%%%%%%%%%%%%%%%%%%%%%%%%%%%%%%

In order to modify the VA approach to appropriately capture the correct instability
growth rates, we choose to modify the reduced VA PDEs so to match the correct
wavenumber cutoffs $k_{c}$.
We thus amend our Lagrangian (\ref{eq6}) by considering the phenomenological
variant:
\begin{eqnarray}
L_y &=& 2X_{t}\left(-AB+u_0^2 \arctan{\frac{B}{A}}\right)-\frac{4}{3}B^3
\notag
\\
&&
\label{eq14}
+\alpha \frac{2B^3}{A^2h(B)}B_y^2+\frac{2\alpha}{3}B^3X_{y}^2,  
\end{eqnarray}
where we have introduced the function
\begin{eqnarray}
 h(B)=B^2-2+2\sqrt{B^4-B^2+1}
\label{eq:h(B)}
\end{eqnarray}
so as to match the wavenumber cutoffs $k_{c}$ obtained from an asymptotic 
approximation to the linear stability problem~\cite{KT}. 
With this modification, the improved equations of motion 
(for the elliptic case $\alpha=-1$) for a DSS yields:
\begin{eqnarray}
A_t&=&-\frac{B^2}{3}X_{yy}+AA_yX_{y},  
\label{eq15}
\\
\label{eq16}
%X_{t}&=&A+\frac{A}{2}X_{y}^2+\frac{A_{yy}}{h}+\frac{AA_y^2}{2B^4}\left[\frac{2u_0^2B^2}{A^2h}-\frac{3B^2}{h}+\frac{B^3}{h^2}h'-\frac{2B^4}{A^2h}\right] \notag \\
X_{t}&=&A+\frac{A}{2}X_{y}^2+\frac{A_{yy}}{h}+\frac{AA_y^2}{2B^2}\left[B\frac{h'}{h^2}-\frac{1}{h}\right].   
\end{eqnarray}
The corresponding linearization for this new reduced PDE model yields:
\begin{eqnarray}
a_t &=& -\frac{u_0^2-\eta^2}{3}\xi_{yy}, \label{eq17} \\
\xi_{t} &=& a+\frac{1}{h(\sqrt{u_0^2-\eta^2})}a_{yy}, \label{eq18}
\end{eqnarray}
which in turn yields the linear dispersion
\begin{eqnarray}
\label{eq:dispVAimproved}
\omega^2&=&\frac{\eta^2-u_0^2}{3}k^2\left[1-\frac{k^2}{h(\sqrt{u_0^2-\eta^2})}\right]. 
\end{eqnarray}
By construction, Eq.~(\ref{eq:dispVAimproved}) captures, as expected, the critical 
wave numbers $k_c^2={h(\sqrt{u_0^2-\eta^2})}$ where instability for $k<k_c$ changes to 
stability for $k>k_c$ for all values of the propagation velocity $A=\eta$, 
see Fig.~\ref{fig:spectrum}. 
Furthermore, as it can be observed from Fig.~\ref{fig:spectrum}, not only the
critical cutoffs $k_c$ match, but the maximum growth rates are also well
approximated with the improved VA equations.
Therefore, we expect that the modified Lagrangian $(\ref{eq14})$, 
leading to the VA equations (\ref{eq15})-(\ref{eq16}),
is able to give a good description for the DSS dynamics for {\em all}
wavenumbers. We will return to this point in Sec.~\ref{sec:numerics},
where we will present our numerical comparisons. 
This is in contrast with the AI approach~\cite{aipaper,aipaper2,aipaper3}
which, by construction, is valid for small wavenumbers and should
thus be expected to be more adequate there (as will be again seen
in Sec.~\ref{sec:numerics}).

%%%%%%%%%%%%%%%%%%%%%%%%%%%%%%%%%%%%%%%%%%%%%%%%%%%%%%%%%%%%%%%%%%%%%%%%%%%%%
\section{Variational Approach for Two Stripes}
\label{sec:2DSS}
%%%%%%%%%%%%%%%%%%%%%%%%%%%%%%%%%%%%%%%%%%%%%%%%%%%%%%%%%%%%%%%%%%%%%%%%%%%%%

In this section we develop a VA approach to follow the interaction
of two DSSs akin to what was obtained using the AI approach in 
Ref.~\cite{aipaper3}.
Consider again the defocusing ($\alpha=-1$) 2D NLS Eq.(\ref{eq:NLS_2D}), %:
%
%\begin{equation}
%\mathrm{i} u_t+\frac{1}{2}u_{xx}+\frac{1}{2}u_{yy}-\left(|u|^2-u_0^2\right)u=0,
%\label{2DSS_eq1}
%\end{equation}
%
which, as mentioned before, may be derived from the renormalized (in the 
$x$-direction) Lagrangian (\ref{eq:L2D})-(\ref{eq5a}), which we now
split as follows:
\begin{eqnarray}
%L&=&\iint_{-\infty}^{\infty} \left[\frac{\mathrm{i}}{2}\left(u^*u_t-uu_t^*\right)\left(1-\frac{u_0^2}{|u|^2}\right)-\frac{1}{2}|u_x|^2-\frac{1}{2}|u_y|^2-\frac{1}{2}\left(|u|^2-u_0^2\right)^2\right]dxdy \notag \\
L = \int_{-\infty}^{\infty}L_y\,dy = \int_{-\infty}^{\infty}\left(L_1+L_2+L_3\right) dy, 
\label{2DSS_eq2}
%-U\left(x\right)|u|^2
\end{eqnarray}
where
\begin{eqnarray}
L_1&=&\frac{\mathrm{i}}{2}\int_{-\infty} ^{\infty} \left(u^*u_t-uu_t^*\right)\left(1-\frac{u_0^2}{|u|^2}\right) dx, \label{2DSS_eq2a} \\
L_2&=&-\frac{1}{2}\int_{-\infty} ^{\infty}\left(|u_x|^2+\left(|u|^2-u_0^2\right)^2\right)dx, \label{2DSS_eq2b} \\
L_3&=&-\frac{1}{2}\int_{-\infty} ^{\infty}|u_y|^2dx. \label{2DSS_eq2c}
\end{eqnarray}
In order to variationally follow the 1D two-dark soliton extension of 
Eq.~(\ref{eq4}) as two DSSs in 2D, we consider the following 
2D {\em symmetric} extension by allowing each dark soliton to have 
its own position and keep a stationary background ($k=0$): 
\begin{eqnarray}
\label{2DSS_eq4}
u\left(x,y,t\right)&=& B\left(y,t\right)T_1(y,t) T_2(y,t)+\mathrm{i} A\left(y,t\right),
\end{eqnarray}
where we use the following short-hand definitions:
\begin{eqnarray}
\label{eq:defs1}
T_1(y,t)&\equiv&\tanh  z_1\left(y,t\right),
\notag
\\
T_2(y,t)&\equiv&\tanh  z_2\left(y,t\right),
\notag
\\
z_1(y,t)&\equiv&B\left(y,t\right)\left(x-X_1\left(y,t\right)\right),
\notag
\\
\notag
z_2(y,t)&\equiv&B\left(y,t\right)\left(x-X_2\left(y,t\right)\right),
\\
\notag
S_1(y,t)&\equiv&\sech  z_1\left(y,t\right),
\\
\notag
S_2(y,t)&\equiv&\sech  z_2\left(y,t\right),
\end{eqnarray}
where $X_1\left(y,t\right)$ and $X_2\left(y,t\right)$ are the spatio-temporal
locations of the two DSSs with $X_1<X_2$. 
The two-DSS ansatz (\ref{2DSS_eq4}) has an overall background level $B$ and 
velocity $A$ ($A>0$ representing the two DSSs moving outwards and $A<0$ inwards)
where, as for the one DSS case, the relation $A^2+B^2=u_0^2=$constant remains valid.
It is important to mention here that a more elaborate two DSS ansatz with 
independent velocities for each dark soliton does not allow for a tractable
VA approach (i.e., integrals that can be obtained in closed form in the 
Lagrangian).
Therefore, we restrict our attention to the ansatz (\ref{2DSS_eq4}) that
has the drawback of having a common (symmetric) velocity for both
dark solitons so that a solution that is not symmetric, namely 
$\dot{X}_2\not=-\dot{X}_1$ will eventually tend to a symmetric one 
($\dot{X}_2=-\dot{X}_1$) 
as time evolves (see Sec.~\ref{sec:numerics} for more details).

Let us now perform the VA approach using the two DSS ansatz (\ref{2DSS_eq4}).
Letting $\Delta\equiv X_2-X_1>0$ (i.e., the two dark solitons do not overlap,
nor change relative positions) and assuming $B\Delta \gg 1$ (i.e., the
two dark solitons are relatively well separated), we have the
following useful approximations:
\begin{eqnarray}
T_1 T_2     &\approx& 1+T_2 - T_1-L(x)e^{-2B\Delta}, \label{2DSS_eq5} \\
T_1 S_2^2   &\approx& S_2^2-F(x)e^{-2B\Delta}, \label{2DSS_eq6}\\
T_2 S_1^2   &\approx& -S_1^2+G(x)e^{-2B\Delta}, \label{2DSS_eq7} \\
S_1^2 S_2^2 &\approx& H(x)e^{-2B\Delta}, \label{2DSS_eq8}
\end{eqnarray}
where the precise form of the functions $L$, $F$, $G$ and $H$ is not
particularly important at this stage as they all contribute
to the same order ($e^{-2B\Delta}$) and will all be combined
appropriately in the final, explicit results below 
[cf.~Eq.~(\ref{2DSS_eq19}) in what follows].
Using this information, we compute:
\begin{eqnarray}
|u|^2&=&u_0^2-B^2\left(S_1^2+S_2^2-S_1^2 S_2^2\right),\label{2DSS_eq9} \\
%\left(|u|^2-u_0^2\right)^2&=&B^4[S_1^4+S_2^4+2S_1^2S_2^2-2S_1^4 S_2^2\notag \\
%&&-2S_1^2S_2^4+S_1^4S_2^4]\notag \\
%&=&B^4[S_1^4+S_2^4+2S_1^2S_2^2-2S_1^4 S_2^2\notag \\
%&&-2S_1^2S_2^4]+B^4h^2e^{-4B\Delta},\label{2DSS_eq10} \\
u_x&=&BD\left(S_1^2T_2+T_1S_2^2\right)\notag \\
%&=&BD\left(-S_1^2 +S_2^2\right)+BD\left(g-f\right)e^{-2D\Delta}, \notag \\
%|u_x|^2 
%&=&B^2D^2\left(S_1^4T_2^2+S_2^4T_1^2+2T_1T_2S_1^2S_2^2\right) \notag \\
%%&=&B^2D^2(S_1^4+S_2^4-S_1^4S_2^2-S_2^4S_1^2\notag \\
%%&&+2T_1T_2S_1^2S_2^2) \label{2DSS_eq11}\\
%%&=&B^2D^2\left(S_1^4+S_2^4-2S_1^2S_2^2\right) \notag \\
%%&& +2B^2D^2\left(-S_1^2+S_2^2\right)\left(g-f\right)e^{-2D\Delta}+B^2D^2\left(g-f\right)^2e^{-4D\Delta}, \notag \\
\frac{\mathrm{i}}{2}\left(u^*u_t-uu_t^*\right)
%&=& \frac{u_0^2}{A}B_tT_1 T_2+AB\left(z_{2t}S_2^2T_1+z_{1t}S_1^2T_2\right)\notag \\
&=&\frac{u_0^2}{A}B_tT_1 T_2+AB\left(z_{2t}S_2^2-z_{1t}S_1^2\right) 
\notag
\\
&&+AB\left(z_{1t}g-z_{2t}f\right)e^{-2D\Delta}.
\label{2DSS_eq12}
\end{eqnarray}
%
%Therefore, from the Lagrangian:
%\begin{eqnarray}
%|u_x|^2+\left(|u|^2-u_0^2\right)^2&=&2B^4S_1^4+2B^4S_2^4-3B^4S_1^4S_2^2-3B^4S_1^2S_2^4 \notag \\
%&&+2B^4T_1T_2S_1^2S_2^2+2B^4S_1^2S_2^2. \label{2DSS_eq13}
%\end{eqnarray}
%
%%We also notice the approximations:
%
%%\begin{eqnarray}
%%\sech z_1 &\approx& 2e^{-D(x-X_1)}, \quad \sech z_2 \approx 2e^{D(x-X_2)}, \quad T_1 \approx 1-2e^{-2D(x-X_1)}\notag \\
% %T_2 &\approx& -1+2e^{2D(x-X_2)}, \quad x \in (X_1,X_2). \notag
%%\end{eqnarray}
%
Using the above we can simplify the integrand of $L_1$ as:
\begin{eqnarray}
&&\frac{\mathrm{i}}{2}\left(u^*u_t-uu_t^*\right)\left(1-\frac{u_0^2}{|u|^2}\right) 
\notag
\\
\approx&&-\frac{u_0^2}{A}B^2B_t\left(\frac{T_1T_2S_1^2}{u_0^2-B^2S_1^2}+\frac{T_1T_2S_2^2}{u_0^2-B^2S_2^2}\right) \notag \notag\\
&&-AB^3\left(z_{2t}\frac{S_2^4}{u_0^2-B^2S_2^2}-z_{1t}\frac{S_1^4}{u_0^2-B^2S_1^2}\right), \label{2DSS_eq15}
\end{eqnarray}
where the approximation,
\begin{eqnarray}
&&\frac{S_1^2 S_2^2-S_1^2-S_2^2}{\frac{u_0^2}{B^2}-S_1^2-S_2^2+S_1^2 S_2^2}
\notag
\\
\approx&&-\frac{S_1^2}{\frac{u_0^2}{B^2}-S_1^2}-\frac{S_2^2}{\frac{u_0^2}{B^2}-S_2^2}+R(x) e^{-2D\Delta}, \notag
\end{eqnarray}
has been used with $z_{1t}=\frac{D_t}{D}z_1-DX_{1t}$ and $z_{2t}=\frac{D_t}{D}z_2-DX_{2t}$.
As it was the case for the functions $L$, $F$, $G$, and $H$ above, 
the precise form of the function $R$ is not relevant at this stage as it will 
be combined in the explicit Lagrangian given in Eq.~(\ref{2DSS_eq19}) below.
We thus get, upon integration,
\begin{eqnarray}
L_1&\approx&2\Delta_t f\left(B\right), \label{2DSS_eq16}
\end{eqnarray}
where
\begin{equation}
f\left(B\right)=-AB+u_0^2 \arctan \frac{B}{A}. \notag \\
\end{equation}
%\begin{eqnarray}
%L&=&\frac{\left[\begin{array}{cc}\frac{u_0^2}{A}B_tT_1T_2+AB\left(z_{2t}S_2^2-z_{1t}S_1^2\right)\\
%+AB\left(z_{1t}g-z_{2t}f\right)e^{-2D\Delta}\end{array}\right]\left(\begin{array}{cc}S_1^2 S_2^2-S_1^2\\
%-S_2^2\end{array}\right)}{\frac{u_0^2}{B^2}-S_1^2-S_2^2+S_1^2 S_2^2} \notag \\
%&&-\frac{B^2}{2}\left(C^2+B^2\right)S_1^4-\frac{B^2}{2}\left(D^2+B^2\right)S_2^4+B^2\left(CD-B^2\right)S_1^2S_2^2 \notag \\
%&&+B^4\left(S_1^4S_2^2+S_1^2S_2^4\right)-B^2\left(D\sech^2z_2-C\sech^2z_1\right)\left(Cg-Df\right)e^{-2D\Delta} \notag \\
%&&-\frac{B^2}{2}\left(Cg-Df\right)^2e^{-4D\Delta}-\frac{B^4}{2}h^2e^{-4D\Delta}, \notag
%\end{eqnarray}
%
On the other hand for the $L_2$ integral, using exact integrations yields:
\begin{equation}
L_2=-\frac{8}{3}B^3+g\left({B\Delta}\right)B^3\approx -\frac{8}{3}B^3+16B^3e^{-2B\Delta}, \label{2DSS_eq14}
\end{equation}
where
\begin{equation}
g\left(z\right)\equiv\frac{-16e^{2z}}{\left(e^{2z}\!-\!1\right)^5}\left[1+\left(9+12z\right)e^{2z}\!-\!\left(9-12z\right)e^{4z}\!-\!e^{6z}\right]. \notag
\end{equation}
Finally, considering the transverse dependence of $u$, from Eq. (\ref{2DSS_eq5}), 
we rewrite (\ref{2DSS_eq4}) as $u=B\left[\tanh D(x-X_2(y,t)) -\tanh D(x-X_1(y,t)) +1\right]+\mathrm{i} A$. 
As in the single dark stripe case, assuming that $B$ and $A$ do not depend on $y$, 
directly but through the relations $D=B$ and $A^2+B^2=u_0^2$ for $D=D(y,t)$,
yields:
\begin{eqnarray}
u_y&=&B\left[\left(D(x-X_2)\right)_yS_2^2-\left(D(x-X_1)\right)_yS_1^2\right]\notag,\\
|u_y|^2&=&B^2\left[z_{2y}^2S_2^4+z_{1y}^2S_1^4-2z_{1y}z_{2y}S_1^2S_2^2\right].\notag
\end{eqnarray}
Assuming small and slow displacements, we can safely neglect the cross terms 
$D_yX_{1y}$ and $D_yX_{2y}$ from $z_{1y}^2$, $z_{2y}^2$ and $z_{1y}z_{2y}$ 
and thus we obtain:
%
%-\frac{1}{6}D_y^2\Delta^3e^{-2D\Delta}+16D^2X_{1y}X_{2y}\Delta e^{-2D\Delta}
\begin{eqnarray}
\frac{L_3}{B^2}&\approx&-\frac{B_y^2}{B^3}\frac{\pi^2-6}{9}-\frac{2}{3}B\left(X_{1y}^2+X_{2y}^2\right)
\notag
\\
&&
+X_{1y}X_{2y}B^3K\left(B\Delta\right), \label{2DSS_eq17}
\end{eqnarray}
where
\begin{eqnarray}
K\left(z\right)&\equiv&4\frac{z\coth z-1}{\sinh^2z}. \notag
\end{eqnarray}

%\begin{eqnarray}
%L_y&=&2\Delta_t\left(-AB+u_0^2 \arctan \frac{B}{A}\right)-\frac{4}{3}\frac{B^2}{D}\left(D^2+B^2\right)+\frac{64}{D}B^4e^{-2D\Delta}+16B^2\left(D^2-B^2\right)\Delta e^{-2D\Delta}\notag \\
%&&-B^2\frac{D_y^2}{D^3}\frac{\pi^2-6}{9}-\frac{2}{3}B^2D\left(X_{1y}^2+X_{2y}^2\right)-\frac{1}{6}B^2D_y^2\Delta^3e^{-2D\Delta}+16B^2D^2X_{1y}X_{2y}\Delta e^{-2D\Delta}. \label{2DSS_eq18}
%\end{eqnarray}
%However at zero-th order in $e^{-2D\Delta}$, the variation in $D$ shows $B=D$.

Now, recalling the introduction of the factor $h(B)$ in Eq.~(\ref{eq:h(B)})
%=B^2-2+2\sqrt{B^4-B^2+1}$ 
to improve the agreement of the growth rates with the numerically
observed ones, we finally obtain the effective Lagrangian:
\begin{eqnarray}
L_y&=&2\Delta_tf\left(B\right)-\frac{8}{3}B^3+g\left(B\Delta\right)B^3
%-K\left(B\right)B_y^2
-\frac{4B^3 B_y^2}{A^2h(B)}
\notag
\\
&&
-\frac{2}{3}B^3\left(X_{1y}^2+X_{2y}^2\right)+X_{1y}X_{2y}B^3K\left(B\Delta \right). 
~~~~
\label{2DSS_eq19}
\end{eqnarray}
%
%where $K(B)\equiv\frac{4B^3}{A^2h(B)}$. 
According to the Euler-Lagrange prescription, taking variations over
$X_1$, $X_2$, and $B$ and recalling that $\Delta\equiv X_2-X_1$
and $A^2+B^2=u_0^2$, yields:
\begin{widetext}
\begin{eqnarray}
A_t\!\!&=&\!\!-\frac{1}{4}B^3g'+\frac{1}{2}AA_y-\frac{1}{6}B^2\Delta_{yy}+\frac{1}{4B}AA_yK\Delta_y+\frac{1}{8}K'AA_y\Delta \Delta_y 
%\notag \\
-\frac{1}{8}K'B^2\Delta_y^2-\frac{1}{8}BK\Delta_{yy}-\frac{1}{4}B^3K'X_{1y}X_{2y}, 
\label{2DSS_eq26} \\
X_{1t}\!\!&=&\!\!-A-\frac{BA}{4}X_1g'+\frac{3}{8}Ag+\frac{AA_y^2}{2B^2h}\left(5-\frac{B}{h}h'\right)-\frac{A_{yy}}{h}
%\notag \\
-\frac{A}{4}\left(X_{1y}^2+X_{2y}^2\right)-\frac{A}{4}X_{1y}X_{2y}X_1BK'+\frac{3}{8}AKX_{1y}X_{2y},  
~~~~~~
\label{2DSS_eq27}\\
X_{2t}\!\!&=&\!\!+A-\frac{BA}{4}X_2g' -\frac{3}{8}Ag-\frac{AA_y^2}{2B^2h}\left(5-\frac{B}{h}h'\right)+\frac{A_{yy}}{h}
%\notag \\
+\frac{A}{4}\left(X_{1y}^2+X_{2y}^2\right)-\frac{A}{4}X_{1y}X_{2y}X_2BK'-\frac{3}{8}AKX_{1y}X_{2y}. 
~~~~~~
\label{2DSS_eq28}
\end{eqnarray}
\end{widetext}
Equations (\ref{2DSS_eq26})--(\ref{2DSS_eq28}) represent one of the main 
results of this work where the VA methodology has been employed to reduce 
the full dynamics of two interacting DSSs in two
spatial dimensions to these three (1+1)D coupled PDEs. 
It is interesting to note that, if one starts with a symmetric initial
configuration $X_2(y,t=0)=-X_1(y,t=0)$, given the symmetry of
Eqs.~(\ref{2DSS_eq27}) and~(\ref{2DSS_eq28}), the dynamics 
preserves this symmetry at all times [i.e., $X_2(y,t)=-X_1(y,t)$ is 
an invariant manifold of the dynamics during the evolution] and hence the
%the dynamics, under the
%restriction $X_2=-X_1$, can be furthered reduced to only two 
coupled PDEs can be reduced to solely the equations
Eq.~(\ref{2DSS_eq27}) for $X_1$ and Eq.~(\ref{2DSS_eq26}) for $A$.

%%%%%%%%%%%%%%%%%%%%%%%%%%%%%%%%%%%%%%%%%%%%%%%%%%%%%%%%%%%%%%%%%%%%%%%%%%%%%
\section{Numerical results}
\label{sec:numerics}
%%%%%%%%%%%%%%%%%%%%%%%%%%%%%%%%%%%%%%%%%%%%%%%%%%%%%%%%%%%%%%%%%%%%%%%%%%%%%

In this section we corroborate that the dynamical reduction, for a single 
DSS and for two interacting DSSs, obtained through the VA approach is 
indeed valid under a wide range of initial conditions.
Moreover, we compare the VA results with the AI
methodology put forward in Refs.~\cite{aipaper,aipaper2,aipaper3} and
showcase where the two display similar results, as well as where the
former represents a significant improvement over the latter.
It is important to stress that, from now on, we use the {\em improved}
VA model that includes the $h(B)$ term introduced in Eq.~(\ref{eq:h(B)}).
Let us start by considering a single DSS. The DSS is always unstable
(snaking instability) to transverse wavenumbers $k$ such that 
$0\leq k \leq k_c$ (see previous section). Therefore, if one considers
a spatial domain $(x,y)\in[-L_x,L_x]\times[-L_y,L_y]$, with a DSS
aligned in the $y$-direction (i.e., using the same notation
as in the previous section), there will always be snaking instability
provided that $L_y>\pi/k_c$. Conversely, if the domain is too small, unstable
transverse modes do not ``fit'' inside the domain and thus the 
DSS is rendered stable (recall for instance how this property can
be used to arrest the instability of DSSs in trapped BECs in
Ref.~\cite{infrared}).

%%%%%%%%%%%%%%%%%%%%%%%%%%%%%%%%%%%%%%%%%%%%%%%%%%%%%%%%%%%%%%%%%%%%%%%
\begin{figure}[tbp]
\begin{center}
\def\myH{4.6cm}
\def\myS{\hspace{0.06cm}}
\includegraphics[height=\myH]{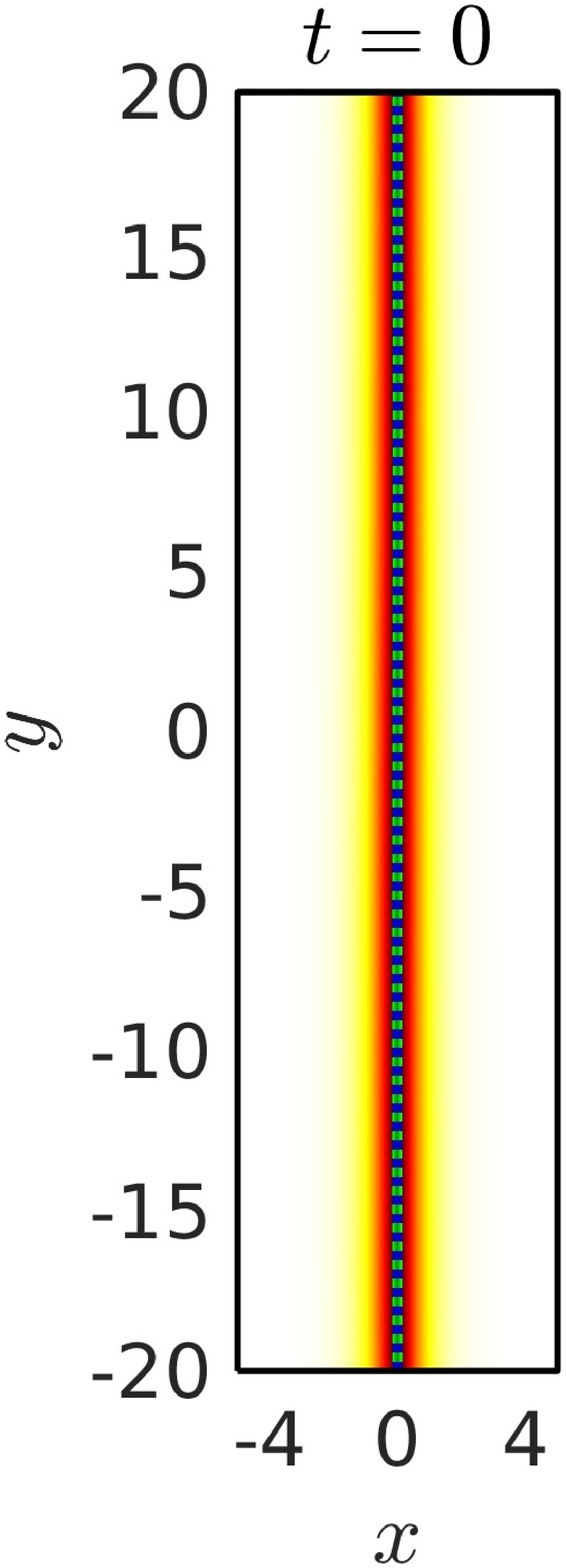}\myS
\includegraphics[height=\myH]{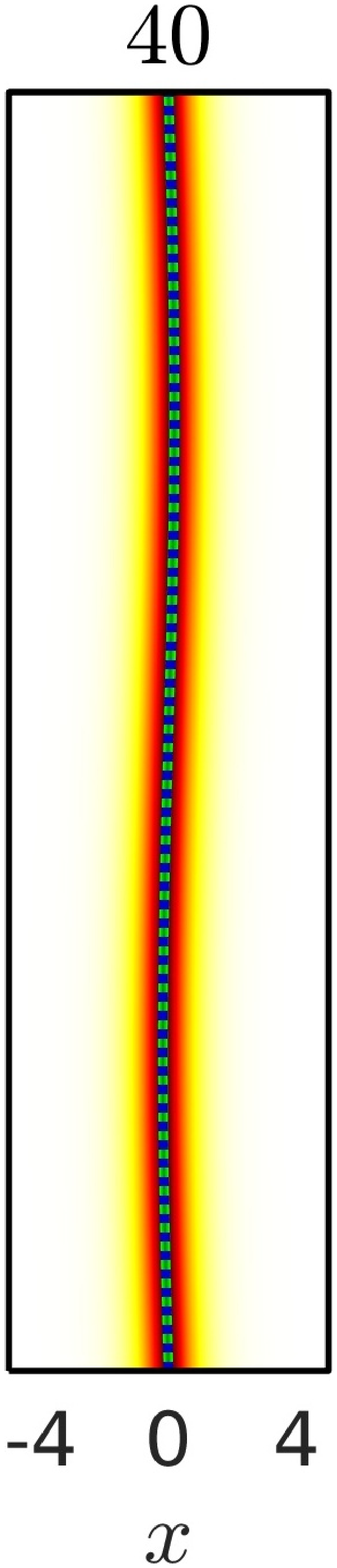}\myS
\includegraphics[height=\myH]{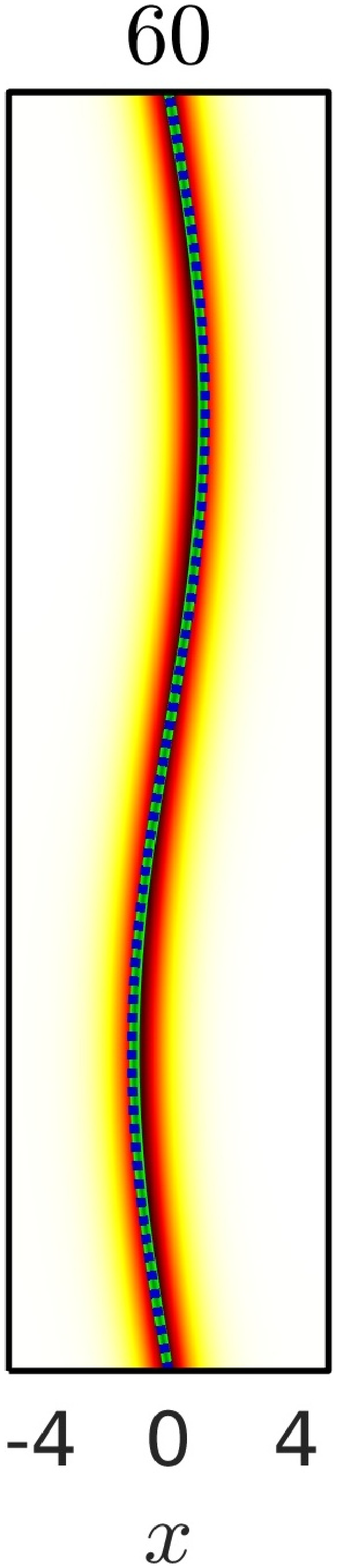}\myS
\includegraphics[height=\myH]{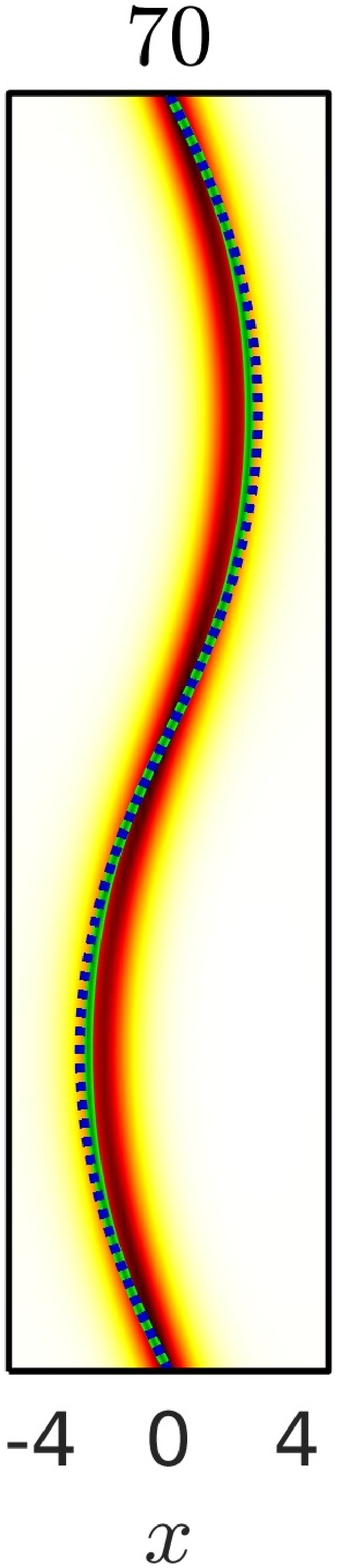}\myS
\includegraphics[height=\myH]{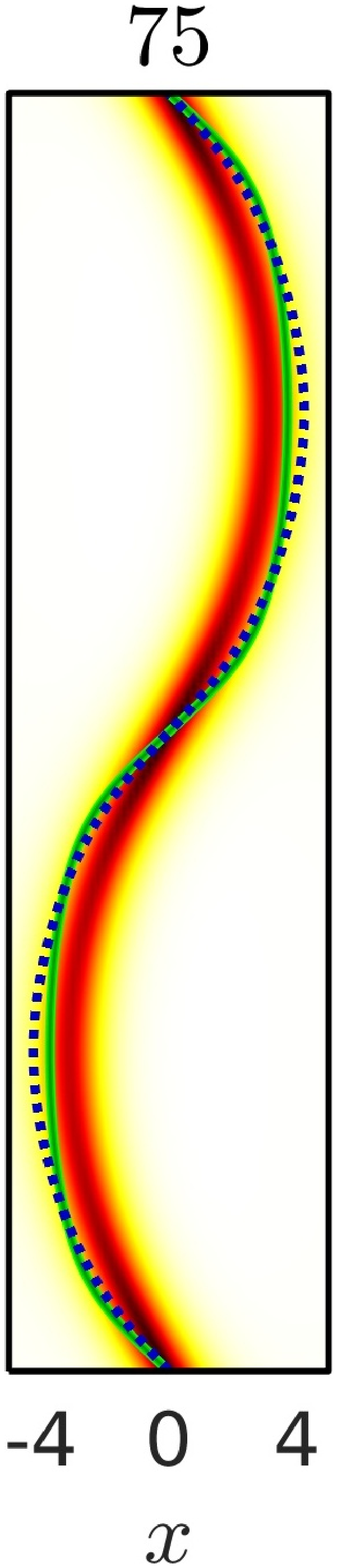}\myS
\includegraphics[height=\myH]{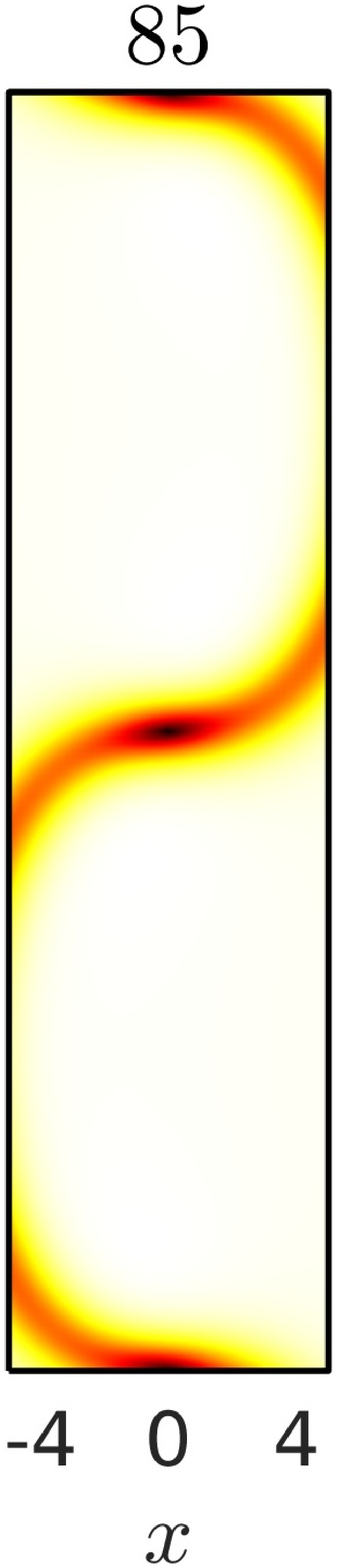}\myS
\includegraphics[height=\myH]{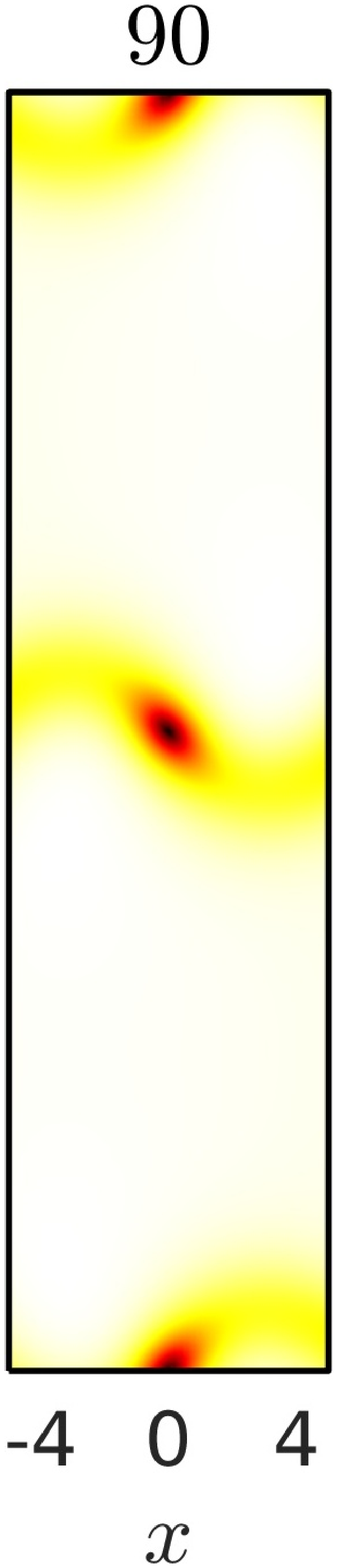}\myS
\includegraphics[height=\myH]{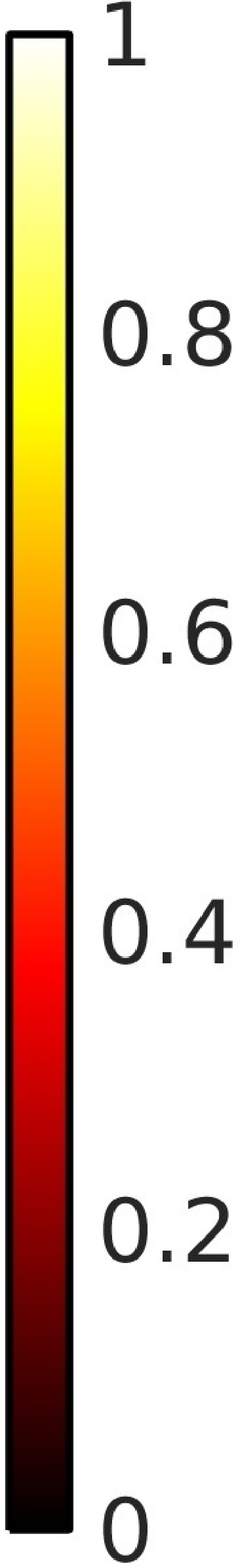}
\caption{(Color online) 
Snaking dynamics of the DSS for $\mu=1$ perturbed by the
first transverse mode of the computational box.
The system is initialized with a stationary DSS perturbed
with the $k=1$ transverse mode  with amplitude 0.01.
Namely, the initial location of the dark soliton at position $y$
is given by $X(y,t=0)=\varepsilon\,\sin(k\pi y/L_y)$
with $\varepsilon=0.01$ and $k=1$.
The modulus of the field, $|u(x,y,t)|$, is plotted in the
$(x,y)$ plane at the times indicated.
The AI prediction is depicted using the dark blue (black) dotted line while 
the corresponding improved VA prediction is depicted by the green (gray) solid line.
The spatial domain is $(x,y)\in[-L_x,L_x]\times[-L_y,L_y]$
with $L_x=L_y=20$ (only the region $-5\leq x\leq 5$
is depicted) under a spatial discretization with $\tt{dx}=\tt{dy}=0.2$.
See Supplemental Material {\tt movie-snake-1} for an animation 
depicting the corresponding dynamics~\cite{SupMat}.
}
\label{fig:snake1}
\end{center}
\end{figure}
%%%%%%%%%%%%%%%%%%%%%%%%%%%%%%%%%%%%%%%%%%%%%%%%%%%%%%%%%%%%%%%%%%%%%%%

%%%%%%%%%%%%%%%%%%%%%%%%%%%%%%%%%%%%%%%%%%%%%%%%%%%%%%%%%%%%%%%%%%%%%%%
\begin{figure}[tbp]
\begin{center}
\def\myH{4.6cm}
\def\myS{\hspace{0.06cm}}
\includegraphics[height=\myH]{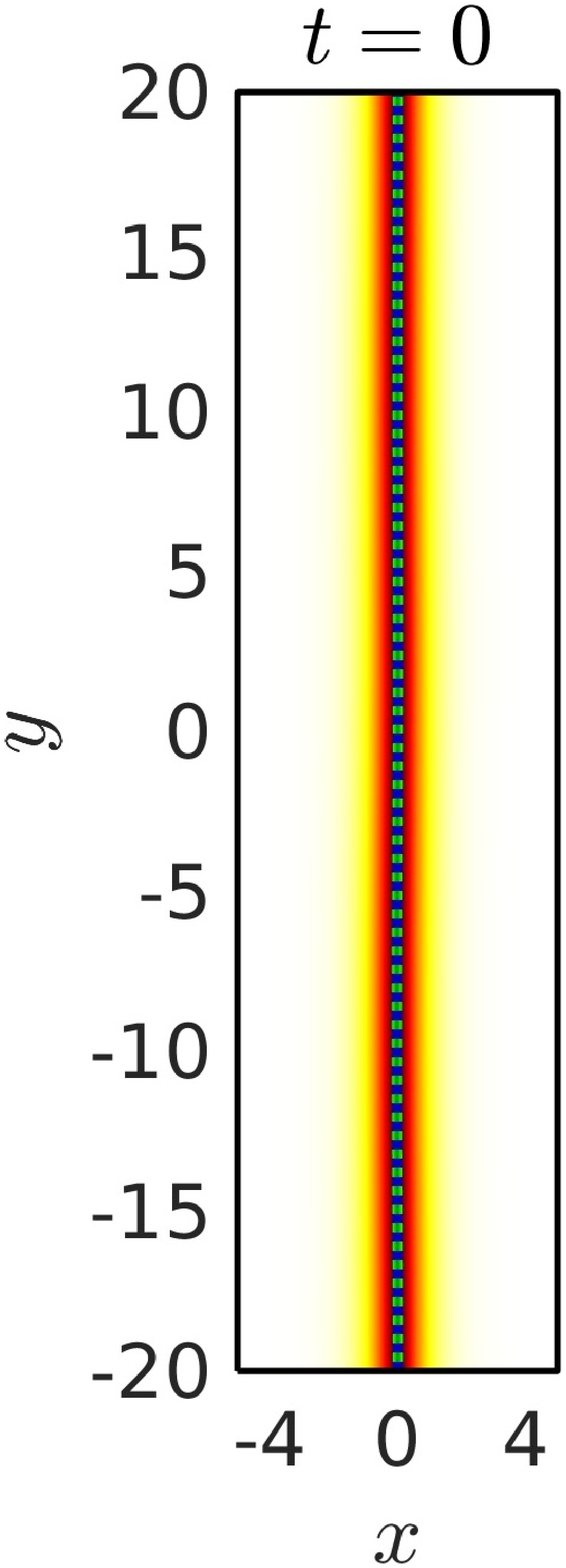}\myS
\includegraphics[height=\myH]{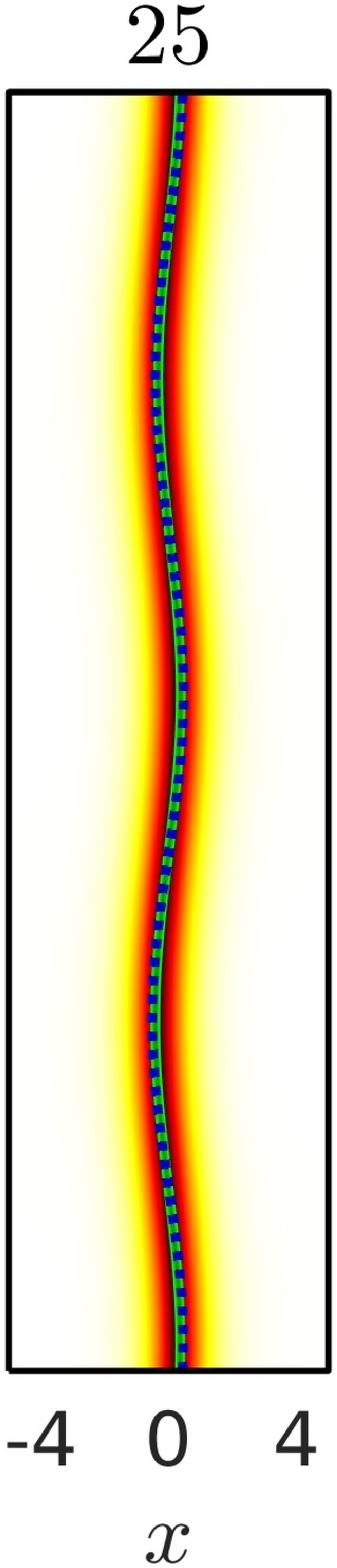}\myS
\includegraphics[height=\myH]{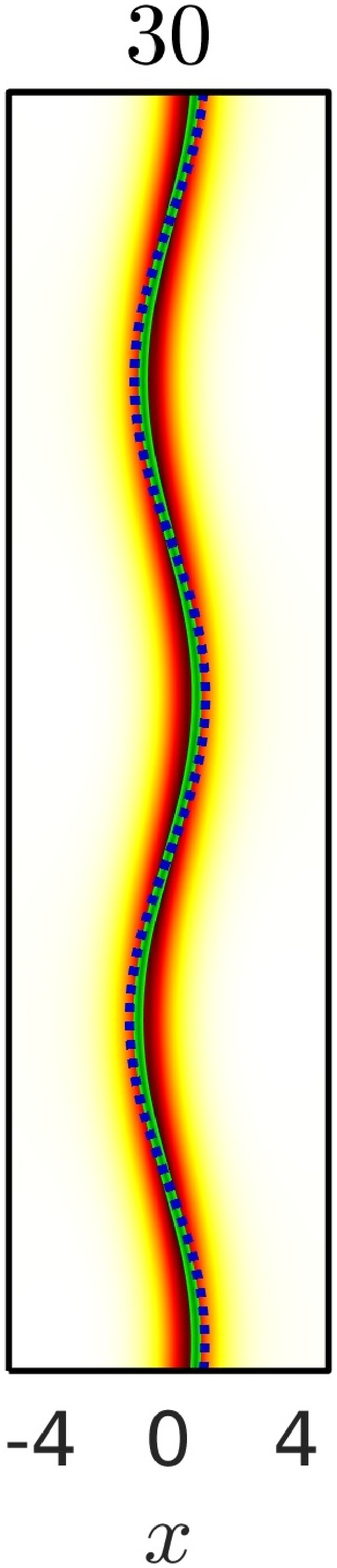}\myS
\includegraphics[height=\myH]{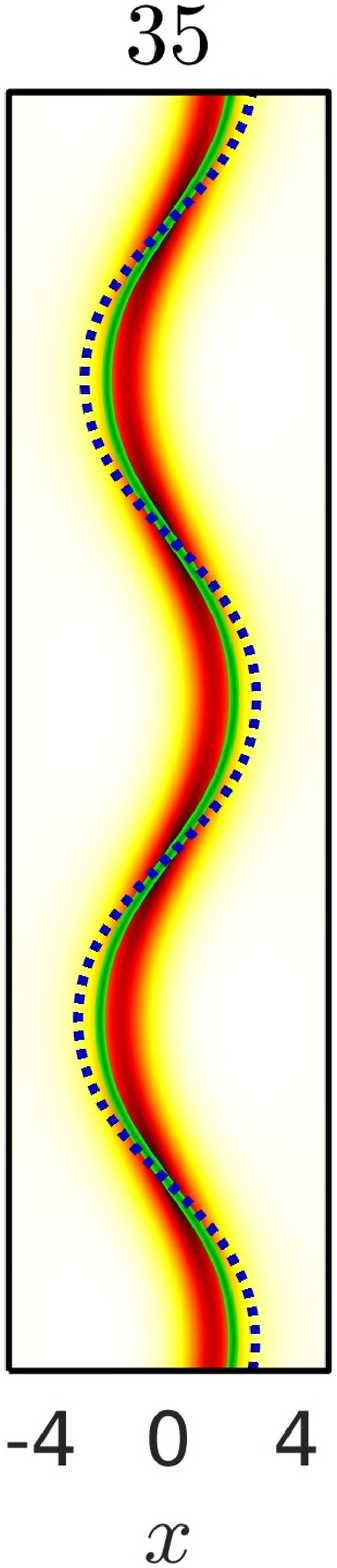}\myS
\includegraphics[height=\myH]{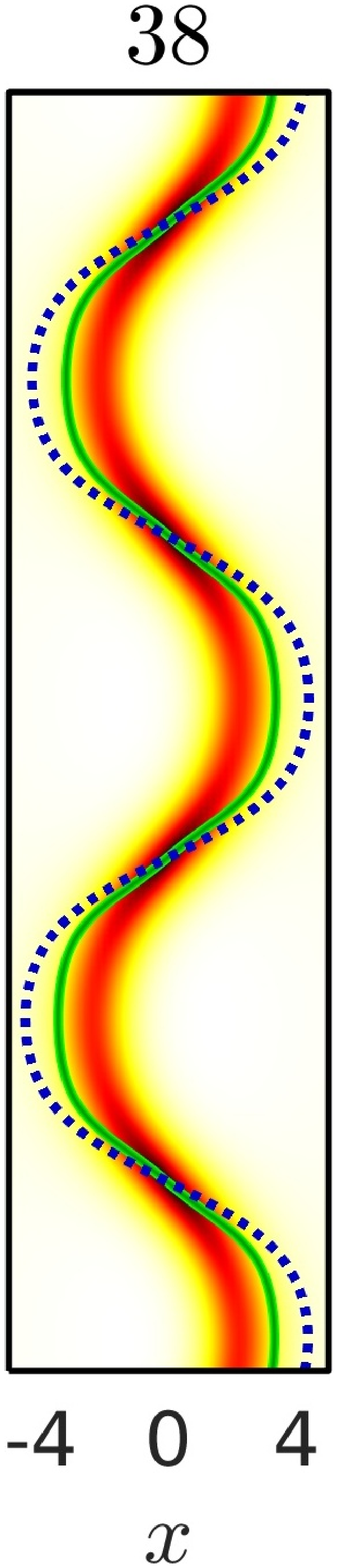}\myS
\includegraphics[height=\myH]{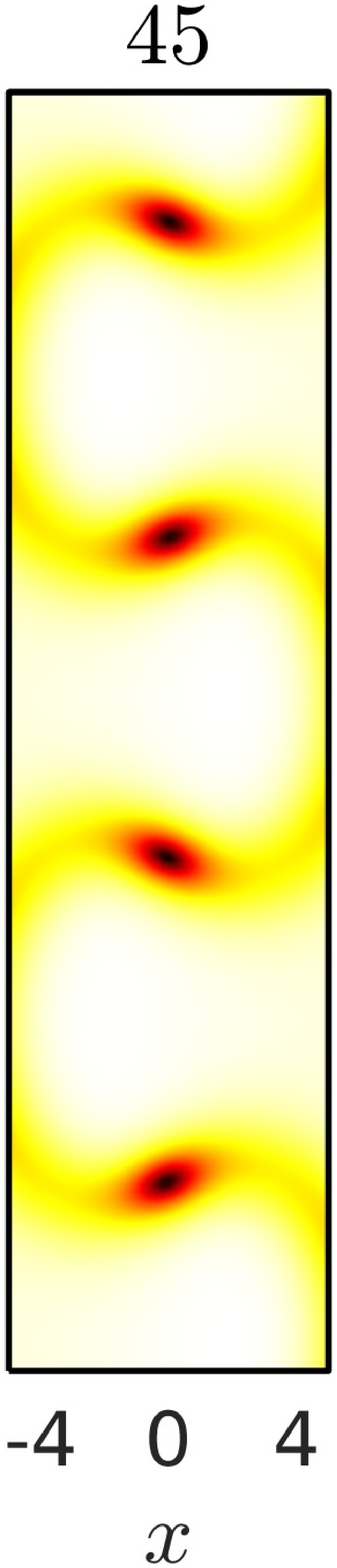}\myS
\includegraphics[height=\myH]{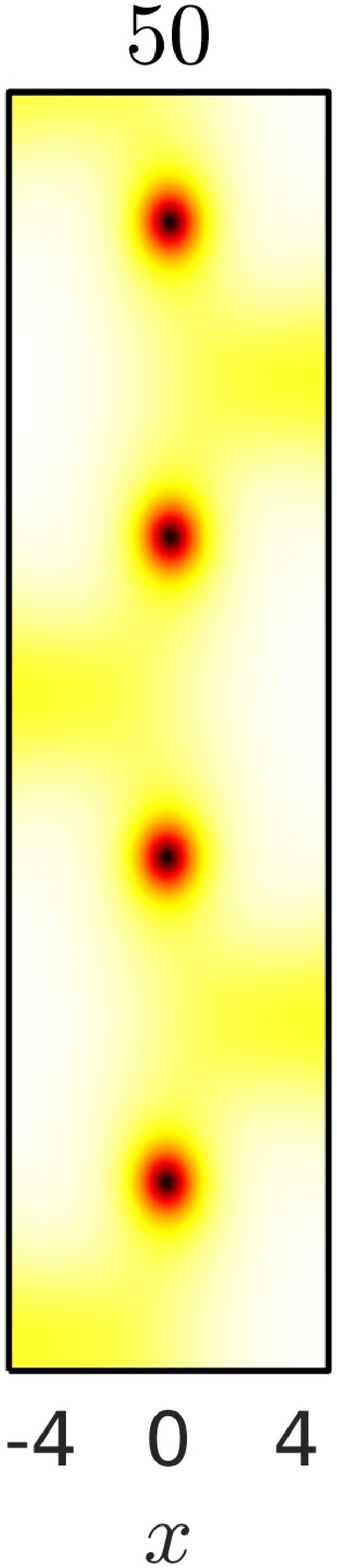}\myS
\includegraphics[height=\myH]{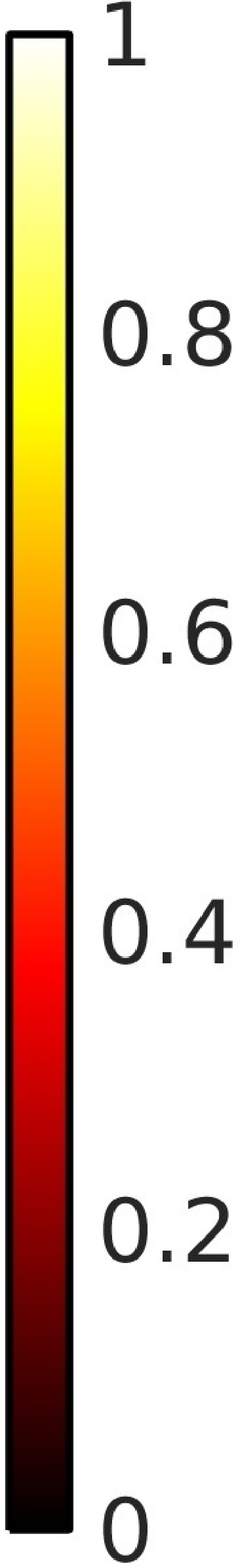}
\caption{(Color online) 
Snaking dynamics of the DSS for $\mu=1$ perturbed by the
second transverse mode.
Same as in Fig.~\ref{fig:snake1} but for $k=2$.
See Supplemental Material {\tt movie-snake-2} for an animation
depicting the corresponding dynamics~\cite{SupMat}.
}
\label{fig:snake2}
\end{center}
\end{figure}
%%%%%%%%%%%%%%%%%%%%%%%%%%%%%%%%%%%%%%%%%%%%%%%%%%%%%%%%%%%%%%%%%%%%%%%

%%%%%%%%%%%%%%%%%%%%%%%%%%%%%%%%%%%%%%%%%%%%%%%%%%%%%%%%%%%%%%%%%%%%%%%
\begin{figure}[tbp]
\begin{center}
\def\myH{6.6cm}
\def\myS{\hspace{0.08cm}}
\includegraphics[height=\myH]{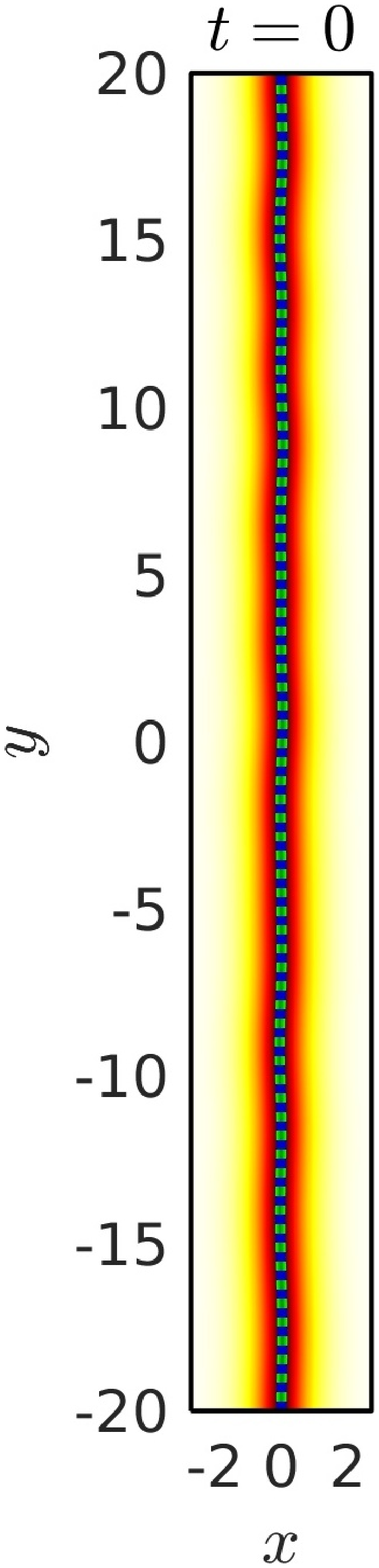}\myS
\includegraphics[height=\myH]{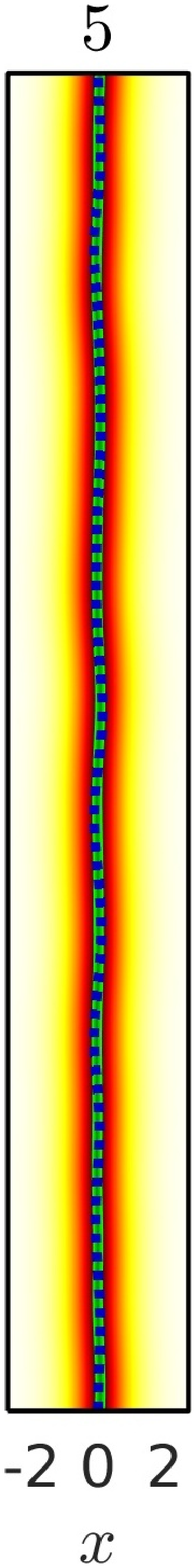}\myS
\includegraphics[height=\myH]{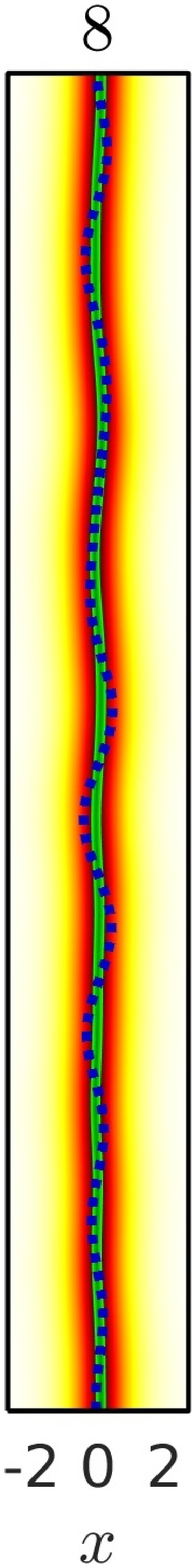}\myS
\includegraphics[height=\myH]{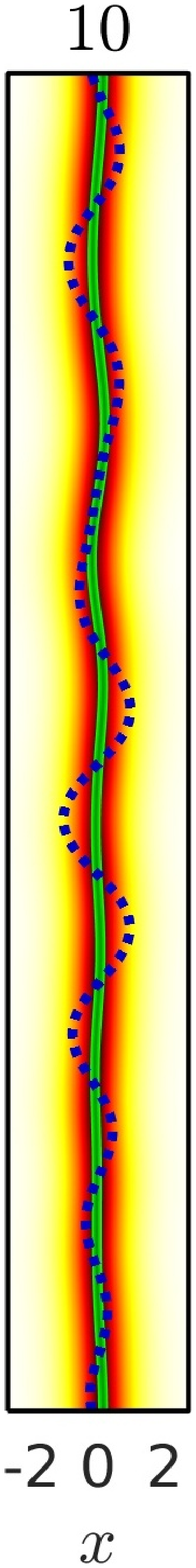}\myS
\includegraphics[height=\myH]{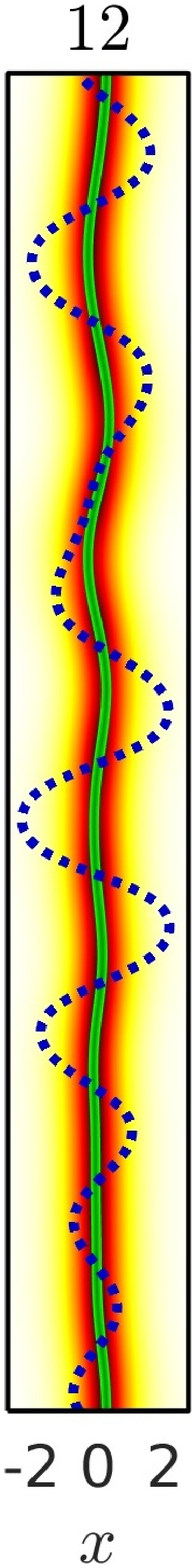}\myS
\includegraphics[height=\myH]{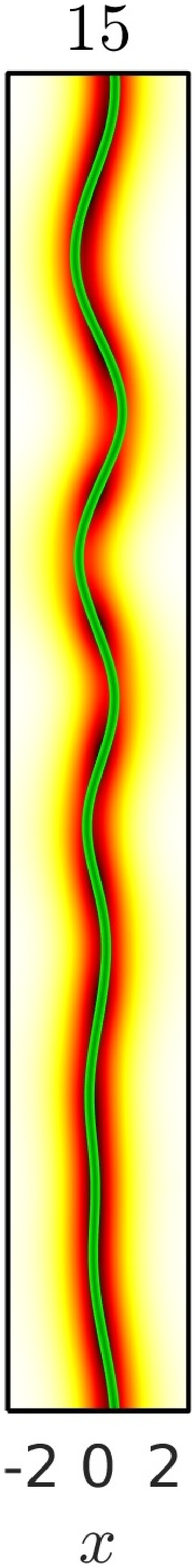}\myS
\includegraphics[height=\myH]{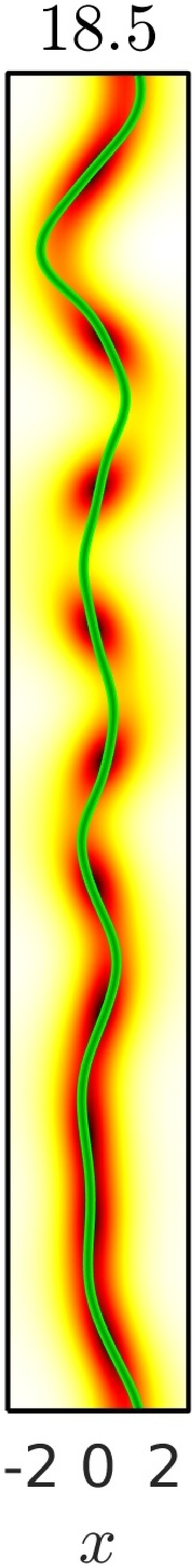}\myS
\includegraphics[height=\myH]{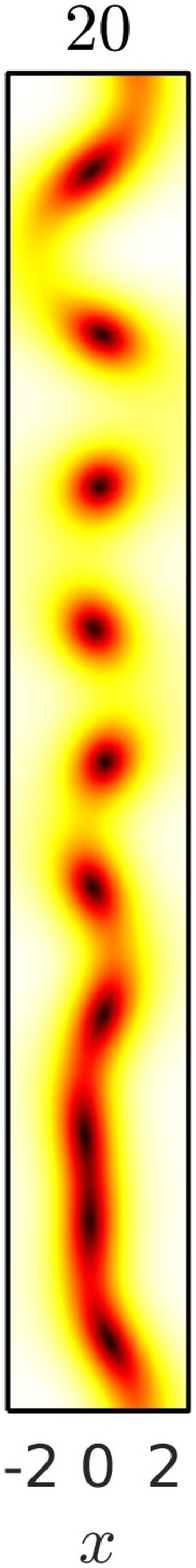}\myS
\includegraphics[height=\myH]{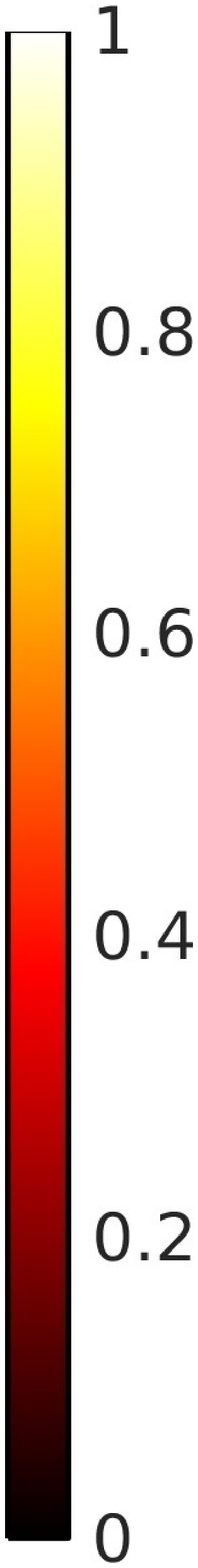}
\caption{(Color online) 
Snaking dynamics of the DSS for $\mu=1$ perturbed by 
a linear combination of the first 10 modes.
Same as in Fig.~\ref{fig:snake1} but for a perturbation of
the initial position of the dark soliton given by
$X(y,t=0)=\sum_{j=1}^{10}\varepsilon_j\,\sin[k\pi(y+\varphi_j)/L_y]$
with $\varepsilon_j=0.01$, $\varphi_j=(j-1)L_y\pi/5$, and $k=j$.
See Supplemental Material {\tt movie-snake-1-10} for an animation
depicting the corresponding dynamics~\cite{SupMat}.
}
\label{fig:snake1-10}
\end{center}
\end{figure}
%%%%%%%%%%%%%%%%%%%%%%%%%%%%%%%%%%%%%%%%%%%%%%%%%%%%%%%%%%%%%%%%%%%%%%%

%%%%%%%%%%%%%%%%%%%%%%%%%%%%%%%%%%%%%%%%%%%%%%%%%%%%%%%%%%%%%%%%%%%%%%%
\begin{figure*}[tbp]
\begin{center}
\def\myH{5.0cm}
\def\myS{\hspace{0.08cm}}
\includegraphics[height=\myH]{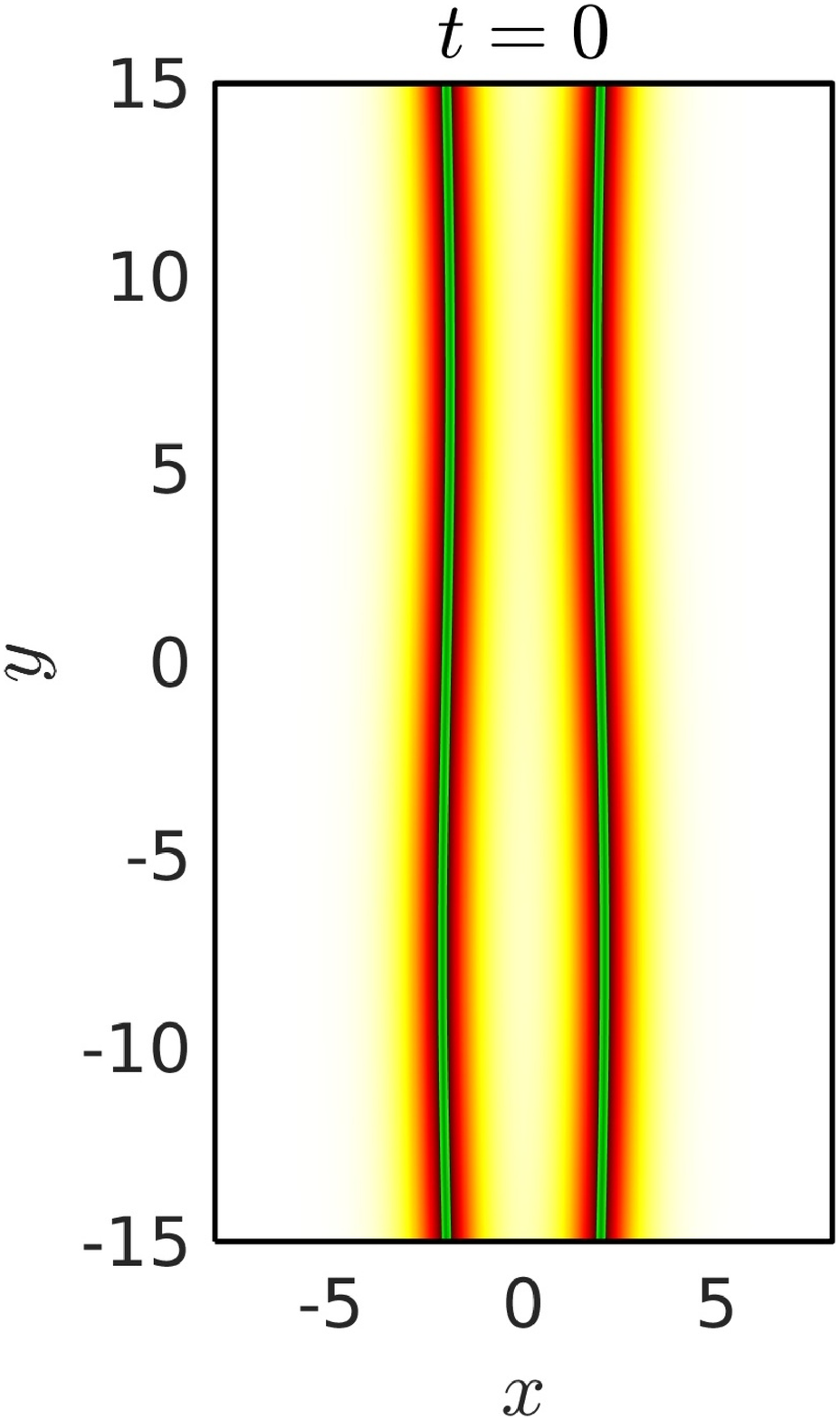}\myS
\includegraphics[height=\myH]{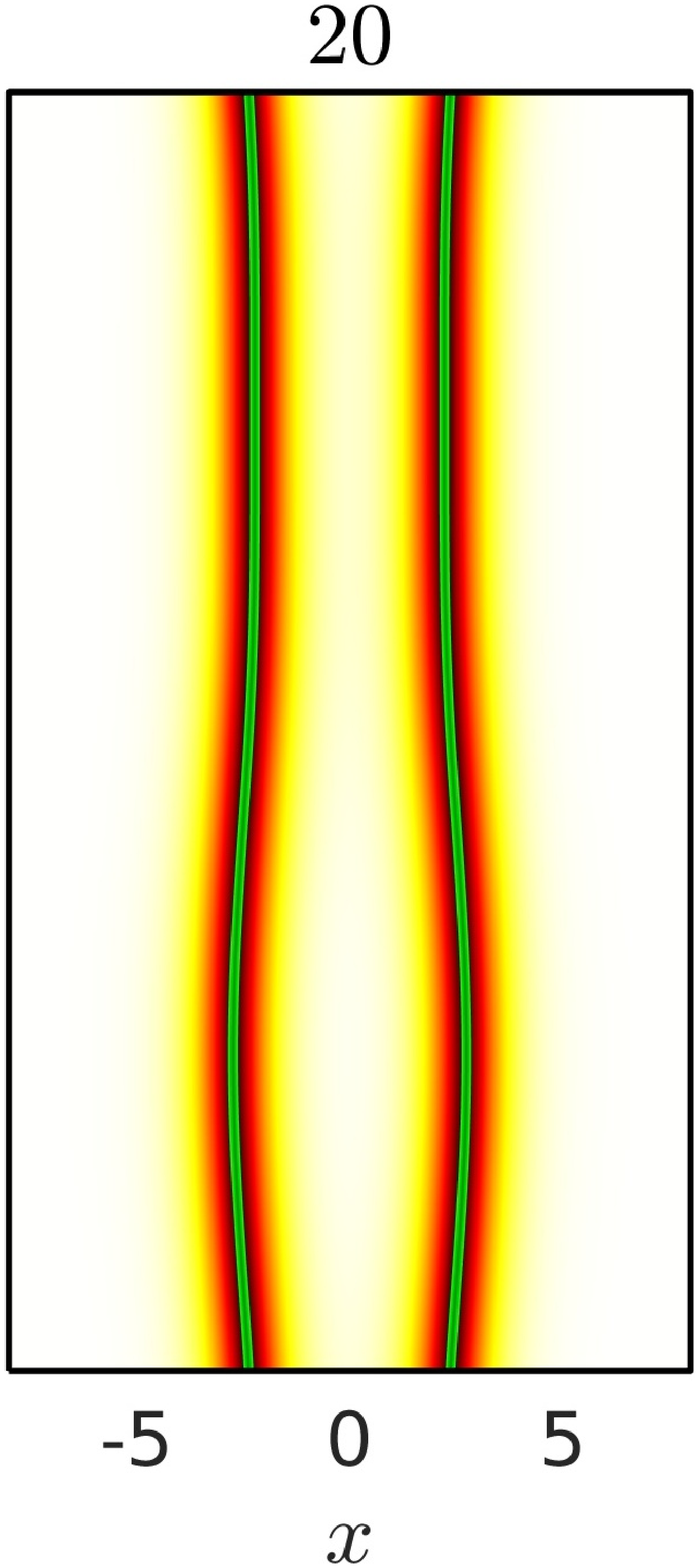}\myS
\includegraphics[height=\myH]{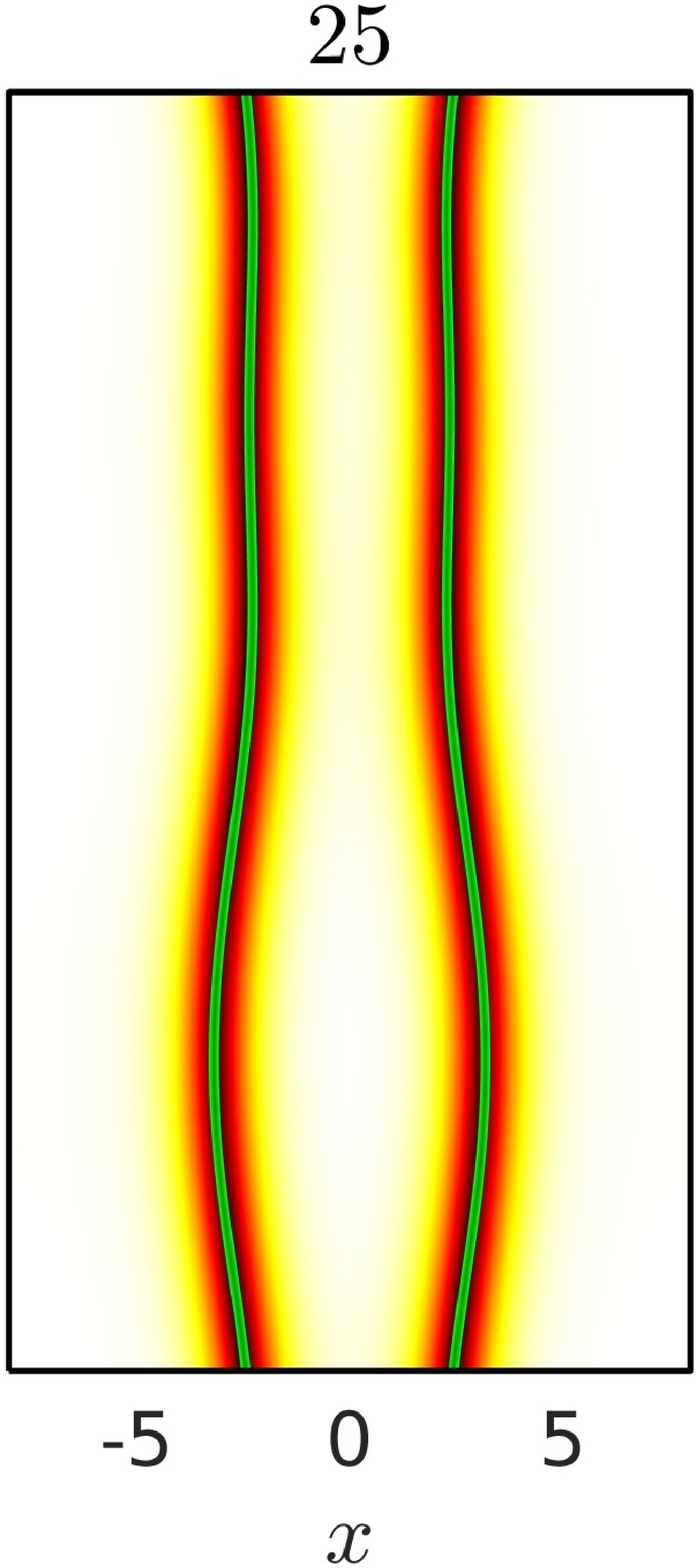}\myS
\includegraphics[height=\myH]{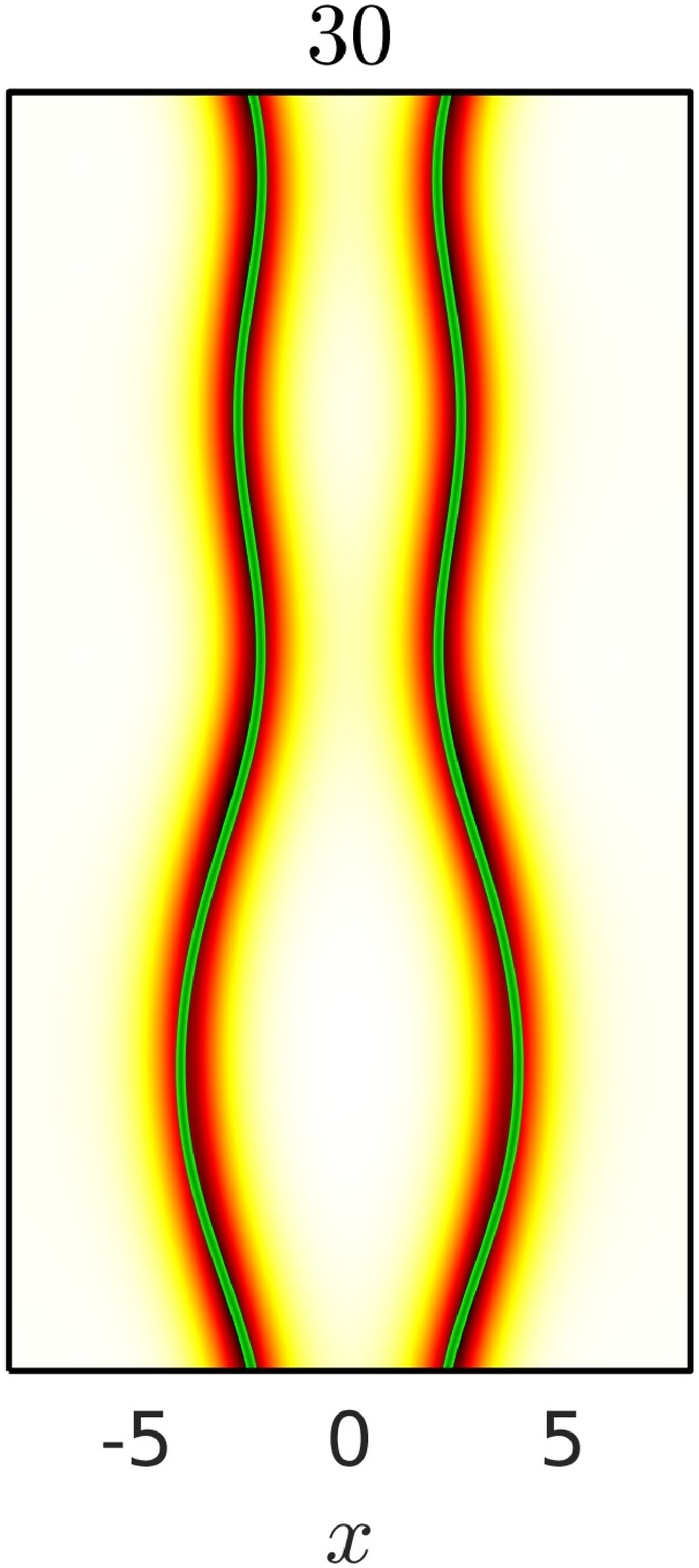}\myS
\includegraphics[height=\myH]{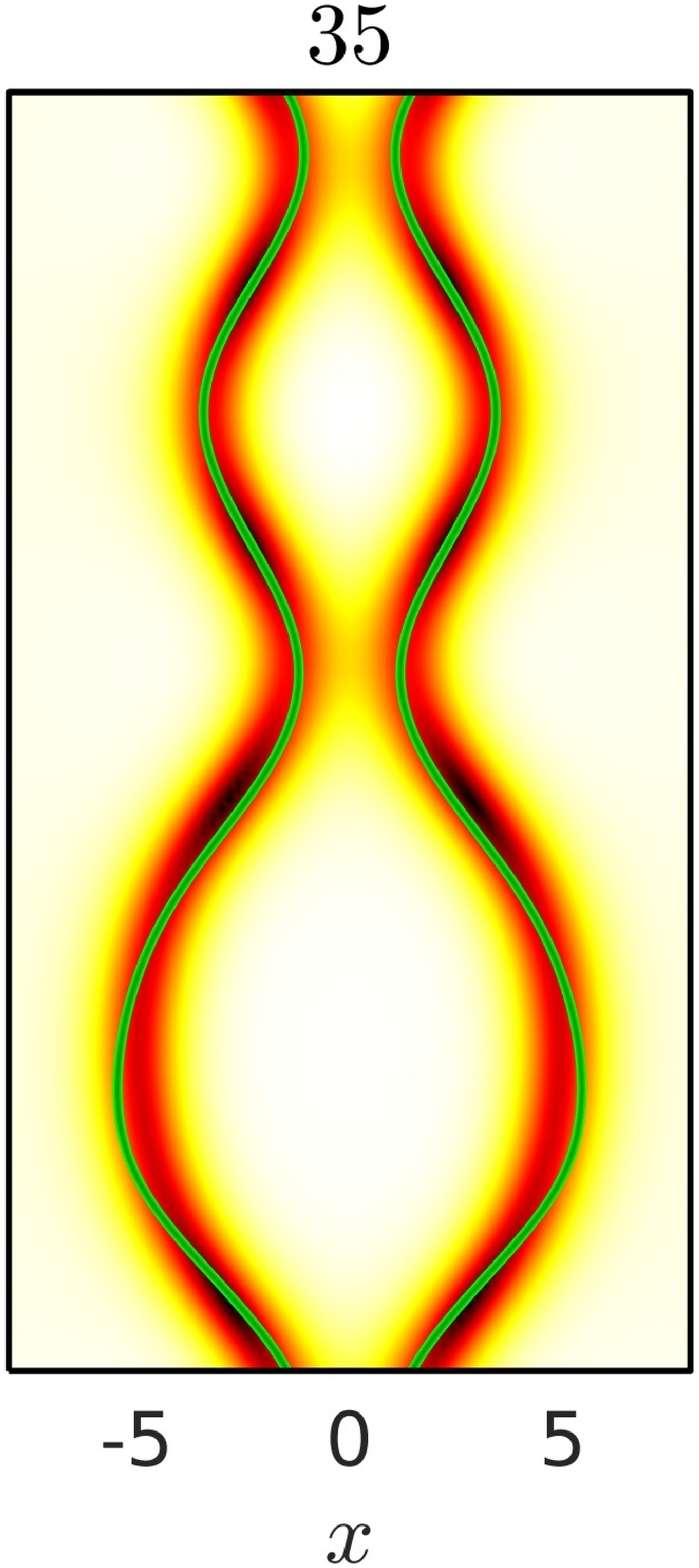}\myS
\includegraphics[height=\myH]{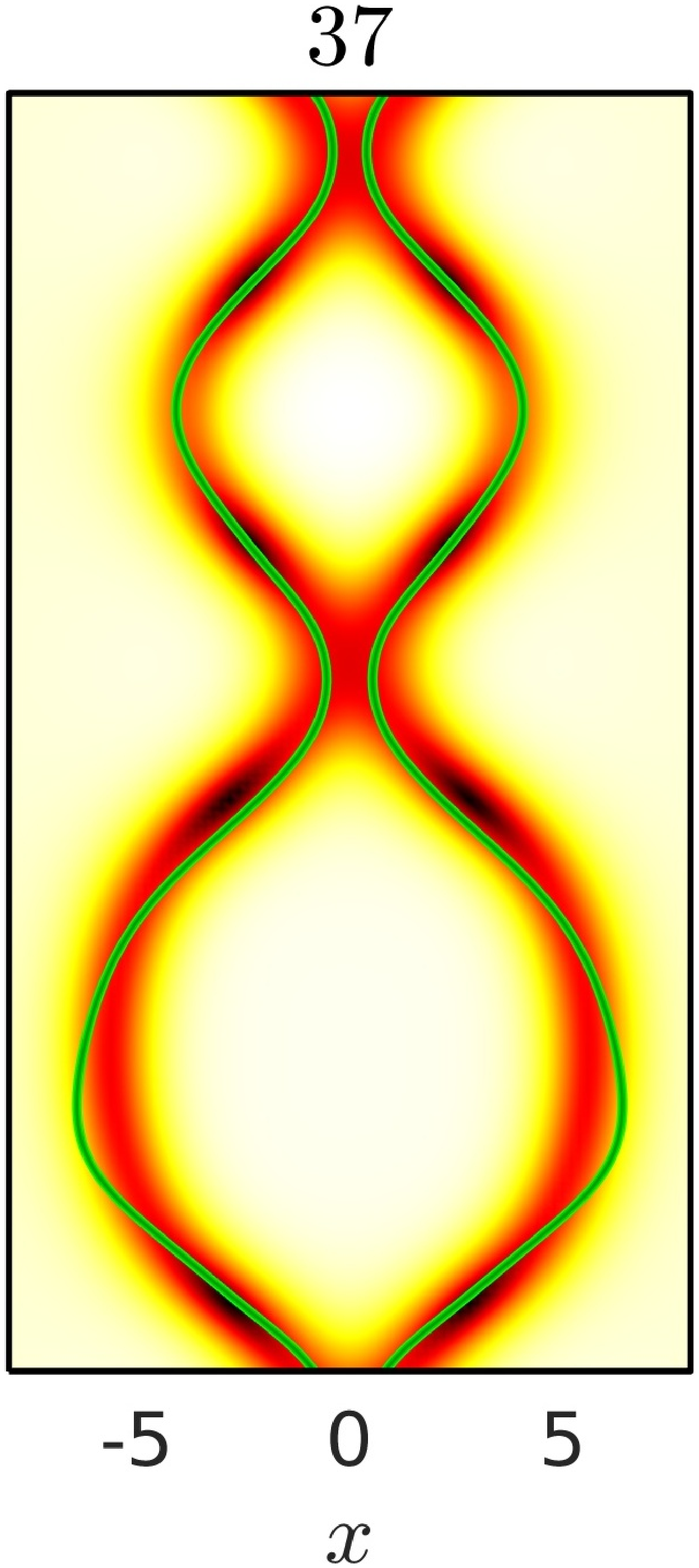}\myS
\includegraphics[height=\myH]{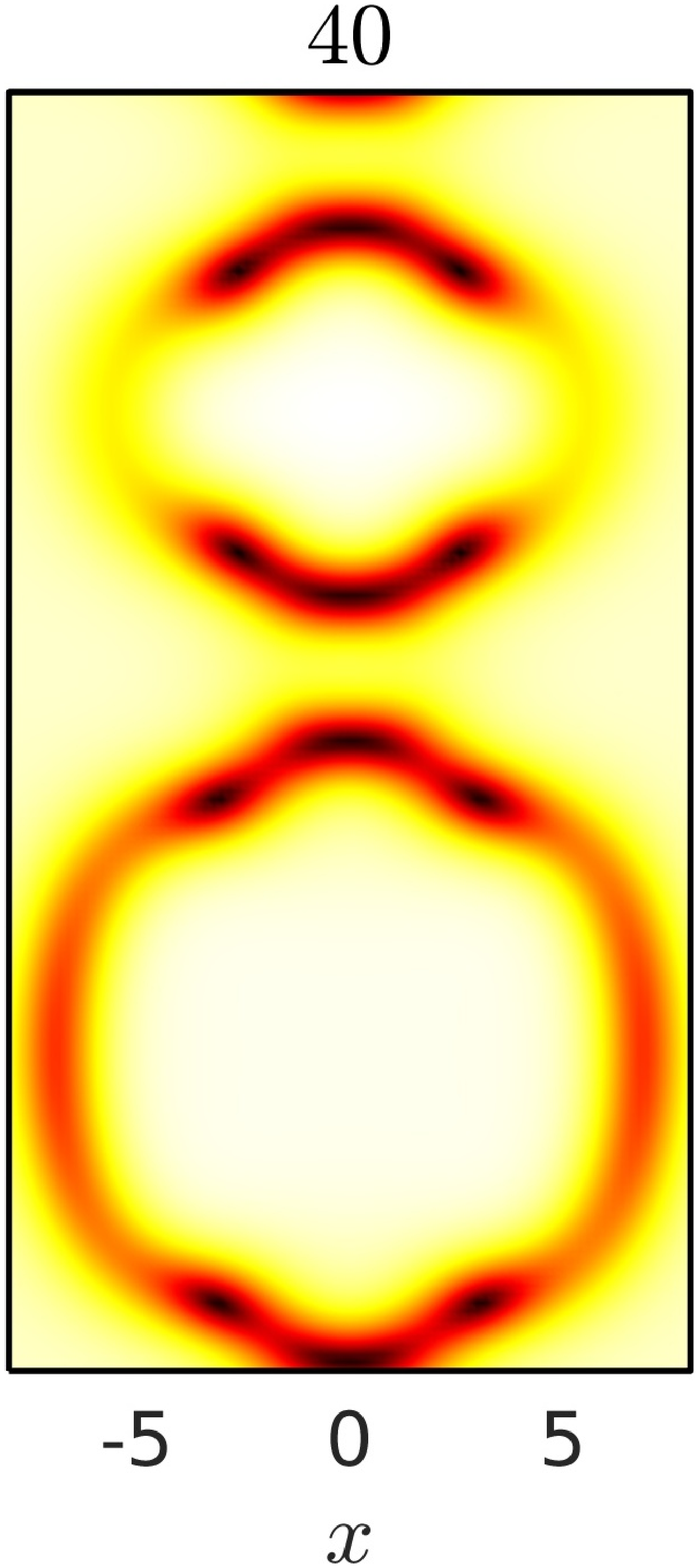}\myS
\includegraphics[height=\myH]{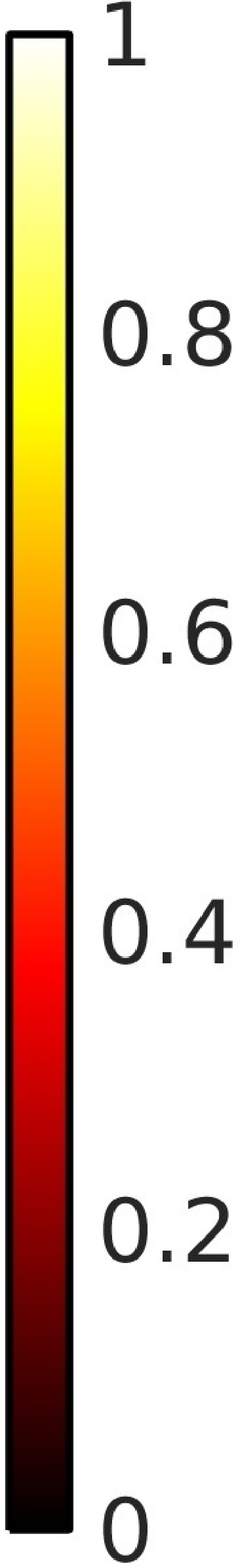}
\caption{(Color online) 
Snaking dynamics of two DSSs for $\mu=1$ perturbed by 
the first transverse mode.
The system is initialized with two DSSs at $x_1=-2=-x_2$
symmetrically perturbed with the $k=1$ transverse mode with 
amplitude 0.1.
Namely, the initial location of each DSSs at position $y$
is given by $-X_1(y,t=0)=X_2(y,t=0)=x_0+\varepsilon\,\sin(k\pi y/L_y)$
with $x_0=2$, $\varepsilon=0.1$ and $k=1$.
The modulus of the field, $|u(x,y,t|$, is plotted in the
$(x,y)$ plane at the times indicated.
The prediction stemming from the improved VA reduced equations 
(\ref{2DSS_eq26})--(\ref{2DSS_eq28}) is depicted by the green solid line.
The spatial domain is $(x,y)\in[-L_x,L_x]\times[-L_y,L_y]$
with $L_x=20$ and $L_y=15$.
See Supplemental Material {\tt movie-2snakes-1} for an animation
depicting the corresponding dynamics~\cite{SupMat}.
}
\label{fig:2snakes1}
\end{center}
\end{figure*}
%%%%%%%%%%%%%%%%%%%%%%%%%%%%%%%%%%%%%%%%%%%%%%%%%%%%%%%%%%%%%%%%%%%%%%%

%%%%%%%%%%%%%%%%%%%%%%%%%%%%%%%%%%%%%%%%%%%%%%%%%%%%%%%%%%%%%%%%%%%%%%%
\begin{figure*}[tbp]
\begin{center}
\def\myH{5.2cm}
\def\myS{\hspace{0.08cm}}
\includegraphics[height=\myH]{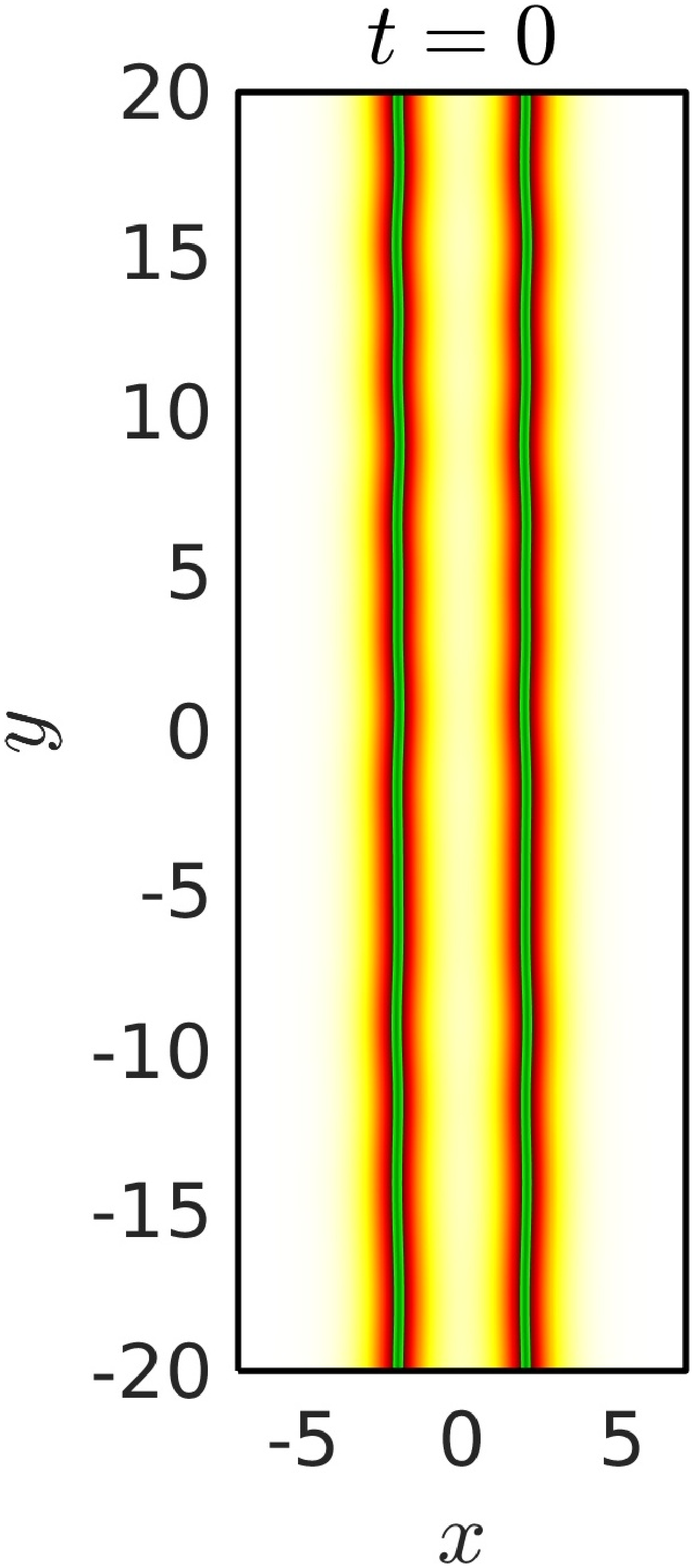}\myS
\includegraphics[height=\myH]{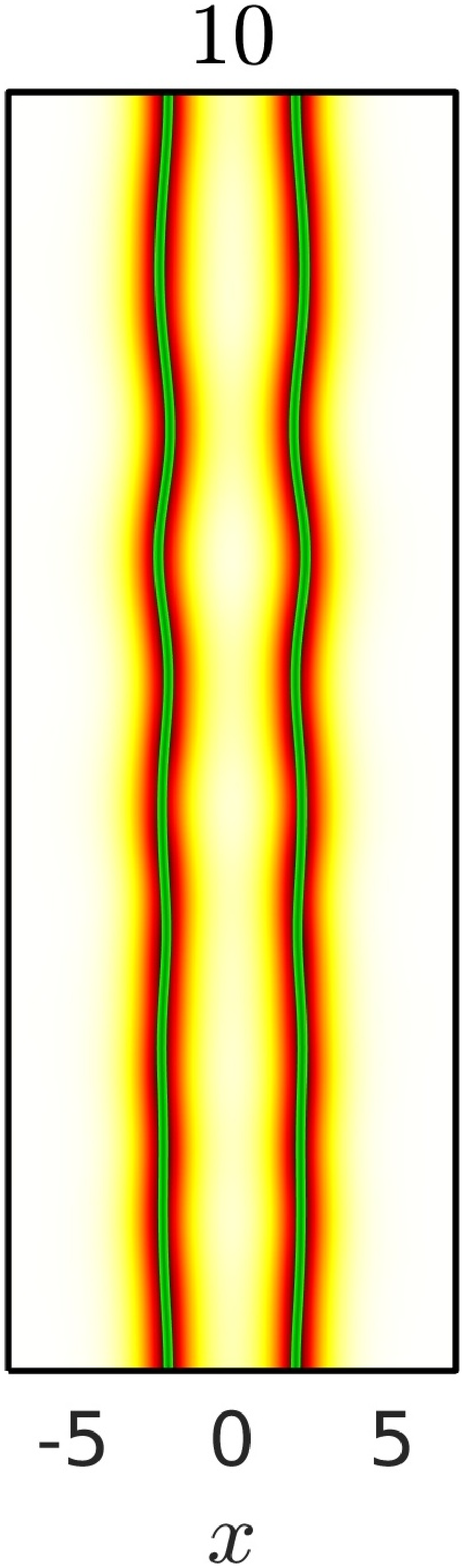}\myS
\includegraphics[height=\myH]{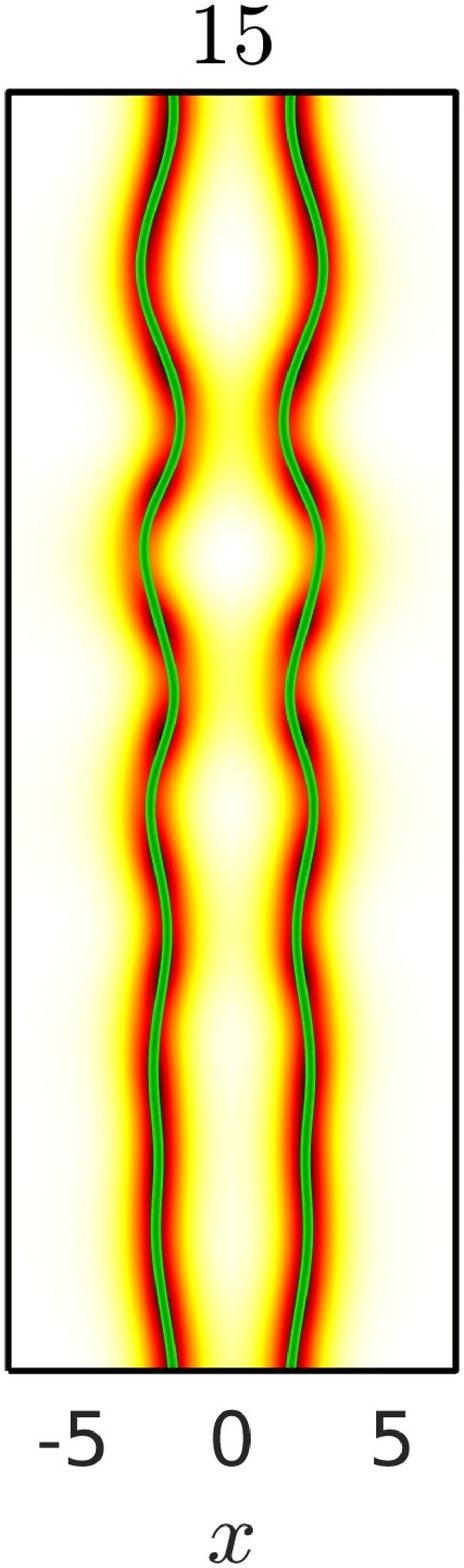}\myS
\includegraphics[height=\myH]{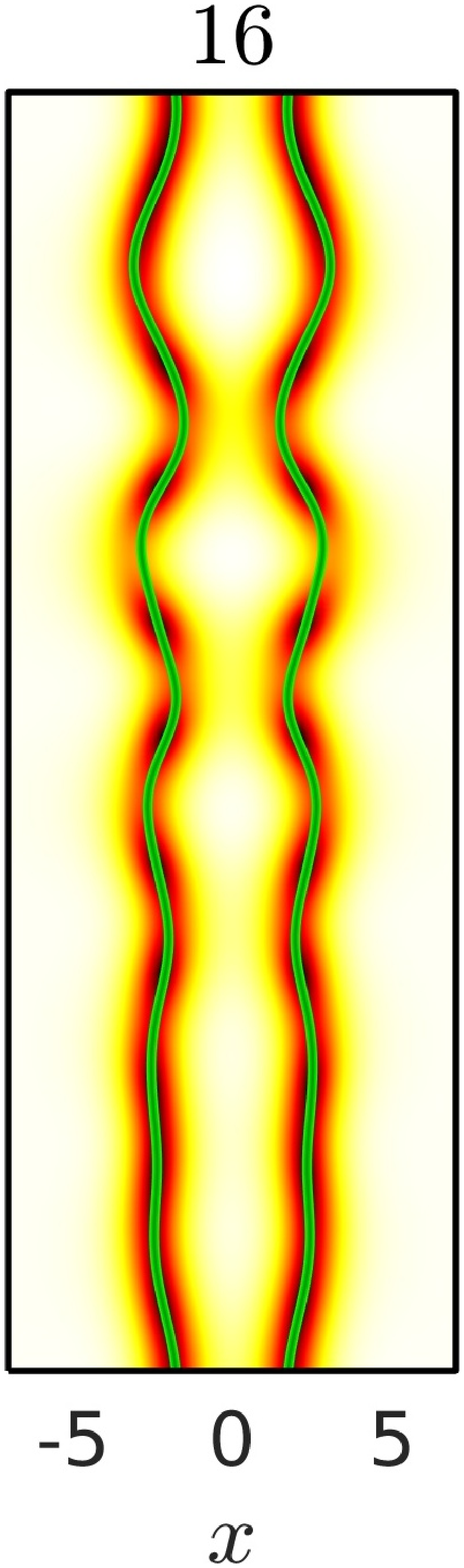}\myS
\includegraphics[height=\myH]{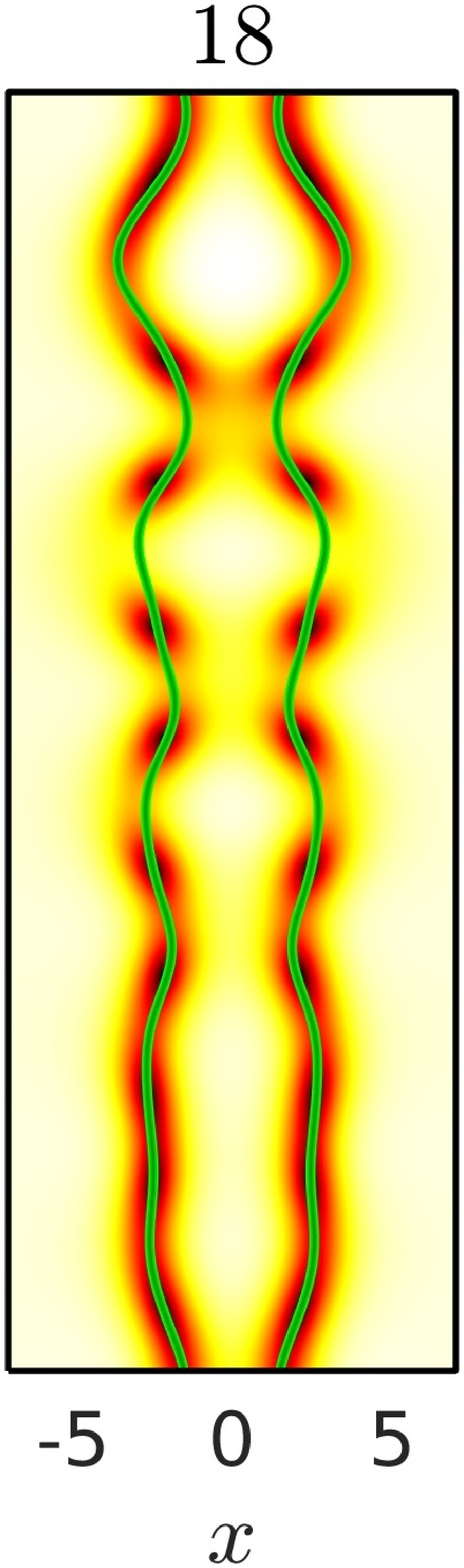}\myS
\includegraphics[height=\myH]{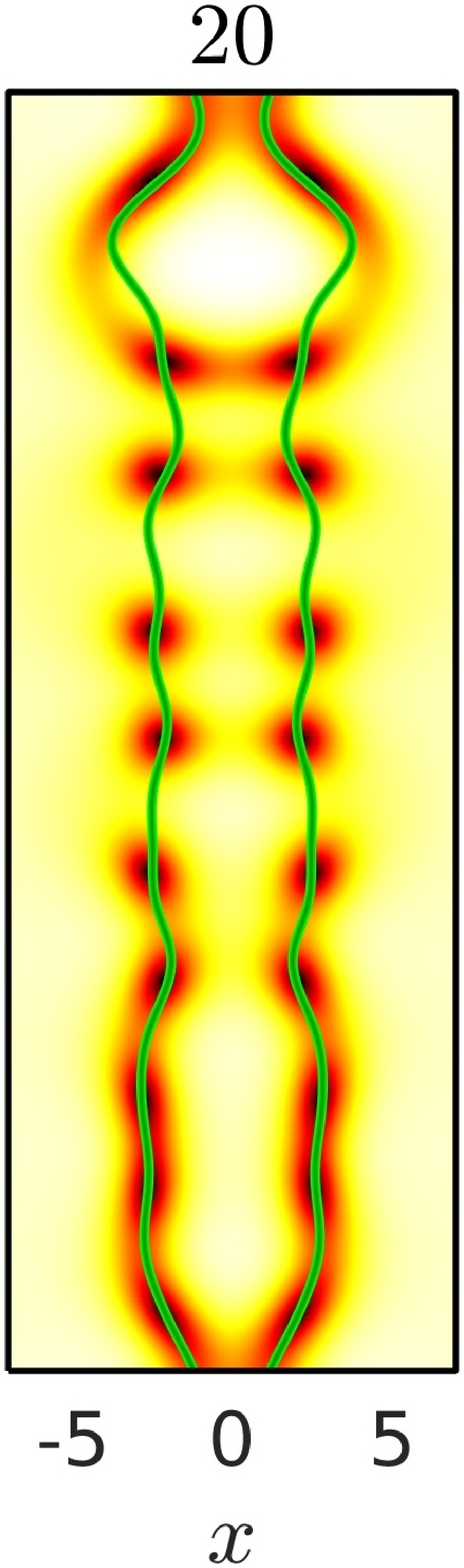}\myS
\includegraphics[height=\myH]{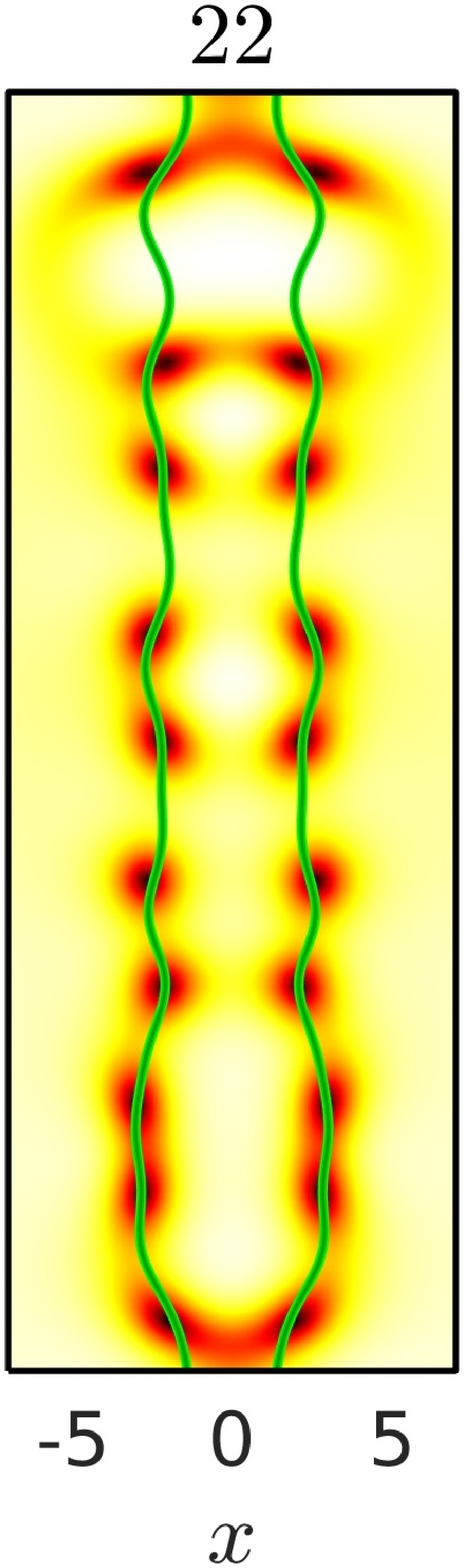}\myS
\includegraphics[height=\myH]{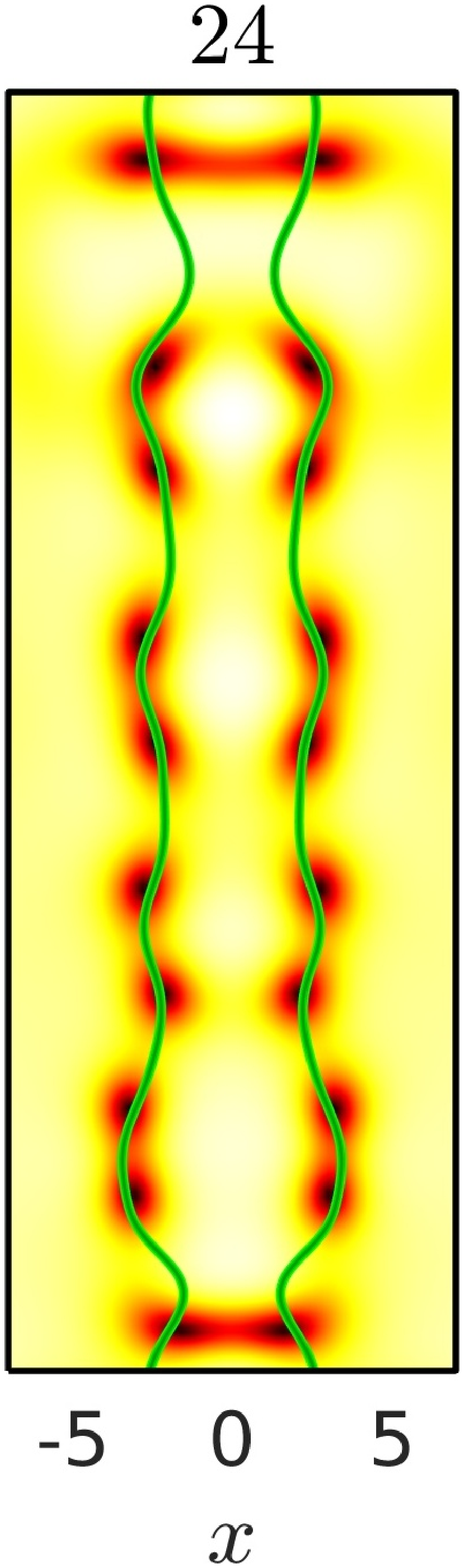}\myS
\includegraphics[height=\myH]{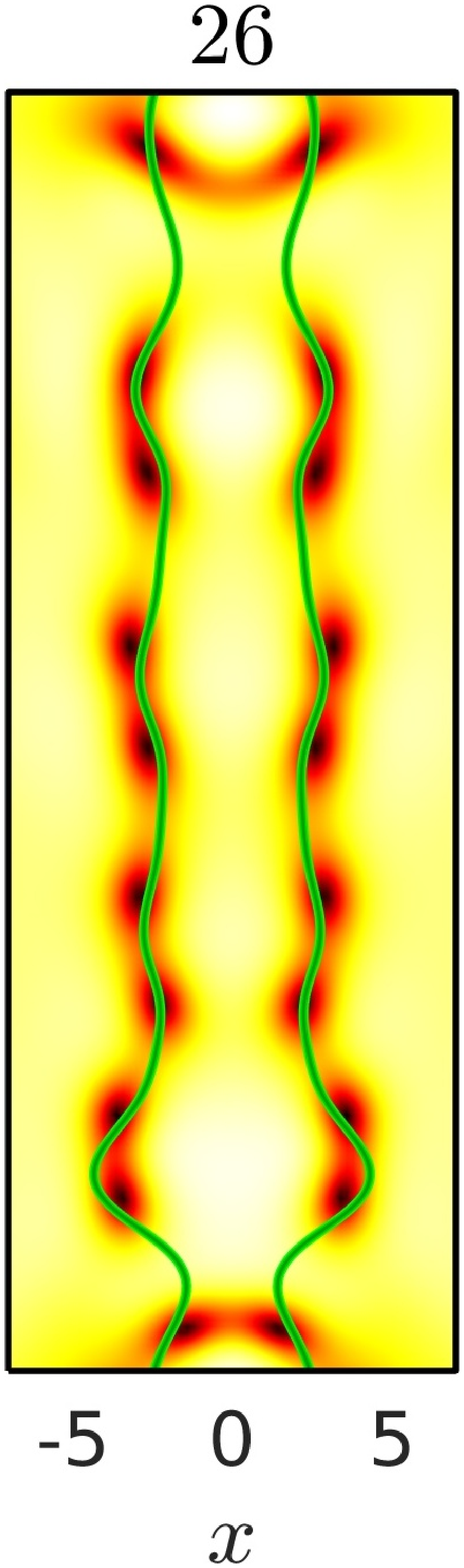}\myS
\includegraphics[height=\myH]{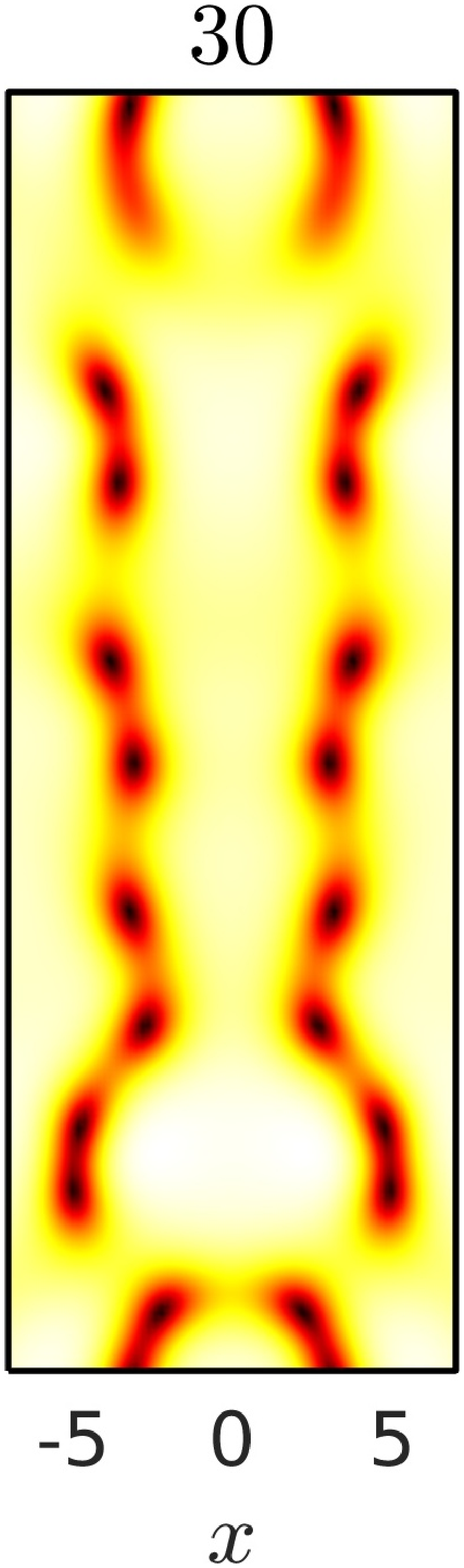}\myS
\includegraphics[height=\myH]{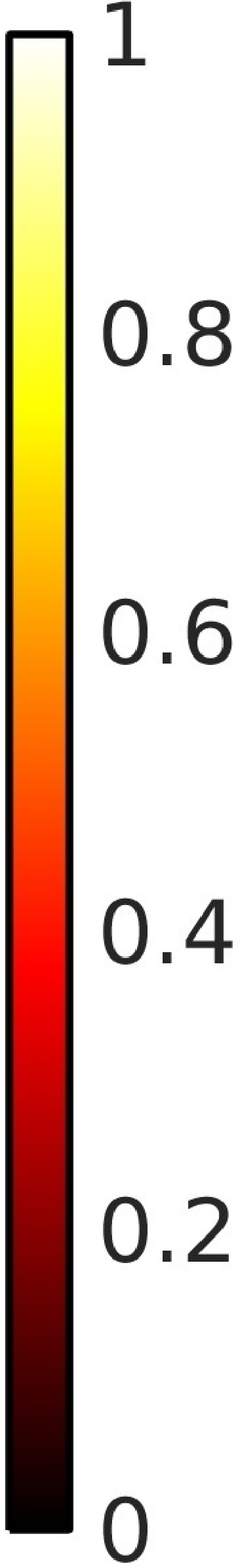}
\caption{(Color online) 
Snaking dynamics of two DSSs for $\mu=1$ perturbed by 
a linear combination of the first 10 modes.
Same as in Fig.~\ref{fig:2snakes1} but for a perturbation of
the initial position of the dark solitons given by
$-X_1(y,t=0)=X_2(y,t=0)=x_0+\sum_{j=1}^{10}\varepsilon_j\,\sin[k\pi(y+\varphi_j)/L_y]$
with $x_0=2$, $\varepsilon_j=0.01$, $\varphi_j=(j-1)L_y\pi/5$, and $k=j$.
The spatial domain is $(x,y)\in[-L_x,L_x]\times[-L_y,L_y]$
with $L_x=L_y=20$.
See Supplemental Material {\tt movie-2snakes-1-10} for an animation 
depicting the corresponding dynamics.
}
\label{fig:2snakes1-10}
\end{center}
\end{figure*}
%%%%%%%%%%%%%%%%%%%%%%%%%%%%%%%%%%%%%%%%%%%%%%%%%%%%%%%%%%%%%%%%%%%%%%%

In order to test the validity of the improved VA approach, we performed a series of 
simulations for the dynamical evolution of a single DSS under different 
perturbations using second order finite differencing in space~\cite{Caplan13a} 
with fourth-order Runge-Kutta with periodic boundary condition along the
$y$-direction and mod-squared Dirichlet (MSD) boundary 
conditions~\cite{Caplan14} along the $x$-direction in order to avoid 
any undesired effects from the boundaries.
The simulations depicted in Figs.~\ref{fig:snake1}--\ref{fig:snake1-10}
are aimed at controllably testing the dynamics of perturbations with
different wavenumbers in the domain $L_x=L_y=20$. In particular, 
Fig.~\ref{fig:snake1} depicts the dynamics ensuing from an
initially stationary [$\eta=A(y,t=0)=0$]
DSS that has been perturbed with the longest possible wavelength
(satisfying the periodic boundary conditions in the $y$-direction).
In addition to testing the VA reduced equations of motion, we also implement 
the AI methodology put forward in Refs.~\cite{aipaper,aipaper2,aipaper3}.
As it can be seen from the figure, both the improved VA [see the thick green (gray) 
curves] and the AI [see the dark blue (black) dotted curves] methodologies give a 
reliable description of the snaking dynamics up to the point where
the DSS loses its transverse dark-soliton-like profile as it nucleates
vortices of alternate signs at the nodes of the perturbation mode.
A few remarks are in order at this point. Firstly, neither the VA nor
the AI methods were designed, by construction, to follow the stripe
after losing its dark-soliton-like stripe shape. Thus, as DSSs
tend to decay into vortex patterns, there will always be a point in
time where the VA (or AI) basic assumptions will be violated
(most notably the ansatz of a dark solitonic stripe being
an accurate descriptor of the full 2D field) and
thus the dynamical reduction will no longer be valid. On the other
hand, as it can be noticed from Fig.~\ref{fig:spectrum}, for low
wavenumbers pertaining the case of Fig.~\ref{fig:snake1}, both VA
and AI are able to appropriately predict the correct growth rate
of perturbations. However, as one notices from Fig.~\ref{fig:spectrum},
both VA and AI tend to slightly {\em overestimate} the growth rates.
This is precisely what it is observed from Fig.~Fig.~\ref{fig:snake1}
where both VA and AI reductions tend to slightly ``run faster'' in
the dynamical destabilization.
Note however, that the VA's overestimation is slightly smaller than the one obtained
through the AI. As we see next, this issue with the AI overestimation will 
become more acute for larger perturbation wavenumbers.

Figure~\ref{fig:snake2} depicts a similar case as the one depicted in
Fig.~\ref{fig:snake1} but for the mode with the second largest wavelength.
Again, both the improved VA and AI tend to slightly overestimate the growth rate
of the perturbation with the VA approximation being better than the
AI one.
In order to test a more complex scenario where the essence
of the improvement of the theory proposed in this work is
most dramatically evident, we perturbed the stationary
DSS with a combination of the first 10 modes. The results are depicted
in Fig.~\ref{fig:snake1-10}.
As it can be observed from the figure, the VA reduction does an
excellent job at following the dynamical stabilization of the DSS.
On the other hand, as we are now introducing larger wavenumbers, the
AI is clearly less accurate (as we expected from its spectral predictions)
and tends to considerably overestimate the growth
rates. In fact, it can be noticed that the VA approximation is even able
to track the full nonlinear dynamics of the DSS up to the point
(and even, arguably, slightly beyond, considering e.g. the snapshot
at $t=18.5$) where it has broken into a pattern including vortex-antivortex pairs.
Furthermore, as the configuration contains larger wavenumbers,
the instability sets in faster and, thus, the slight growth rate
overestimation provided by the VA is now downplayed over the span
of time before the DSS breaks into vortices.
This highlights that in a typical scenario where several unstable
modes are present, the improved VA will be an excellent reduced 
description of the full dynamics of single DSSs.

%%%%%%%%%%%%%%%%%%%%%%%%%%%%%%%%%%%%%%%%%%%%%%%%%%%%%%%%%%%%%%%%%%%%%%%
\begin{figure}[tbp]
\begin{center}
\def\myH{4.65cm}
\def\myS{\hspace{0.06cm}}
\includegraphics[height=\myH]{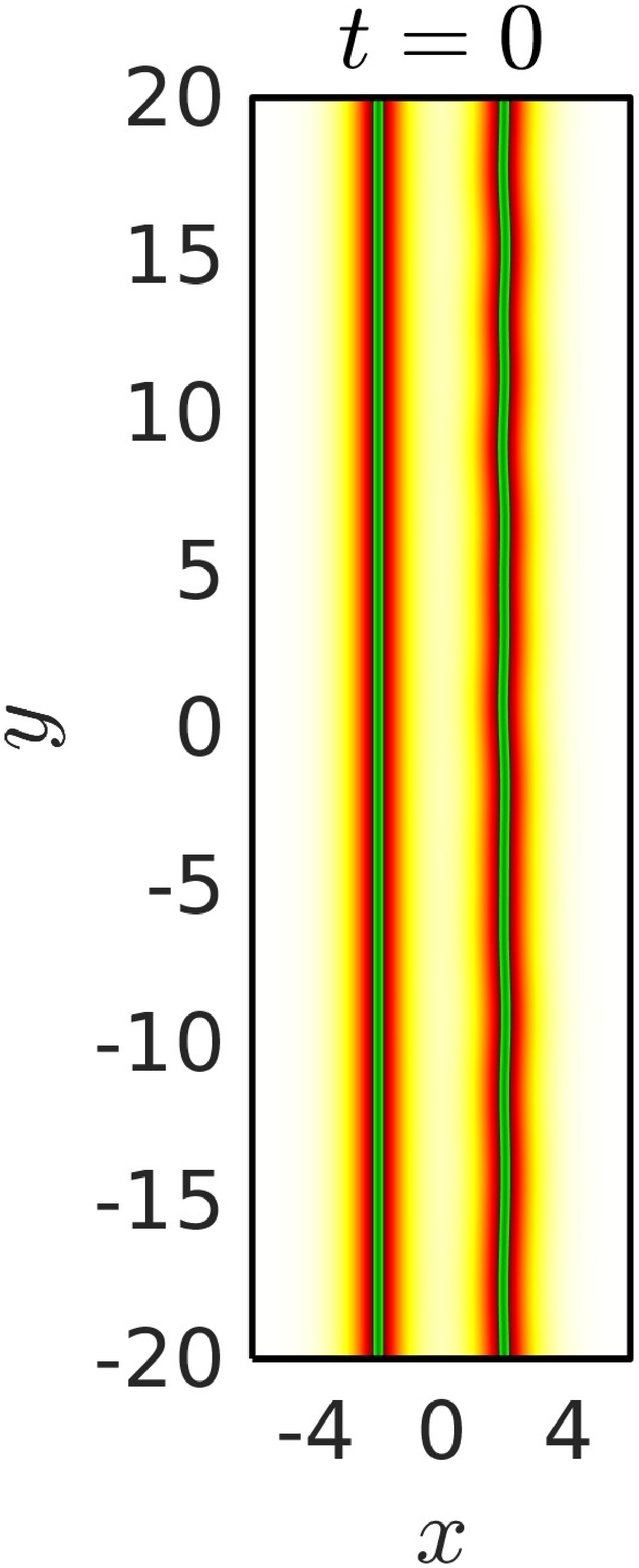}\myS
\includegraphics[height=\myH]{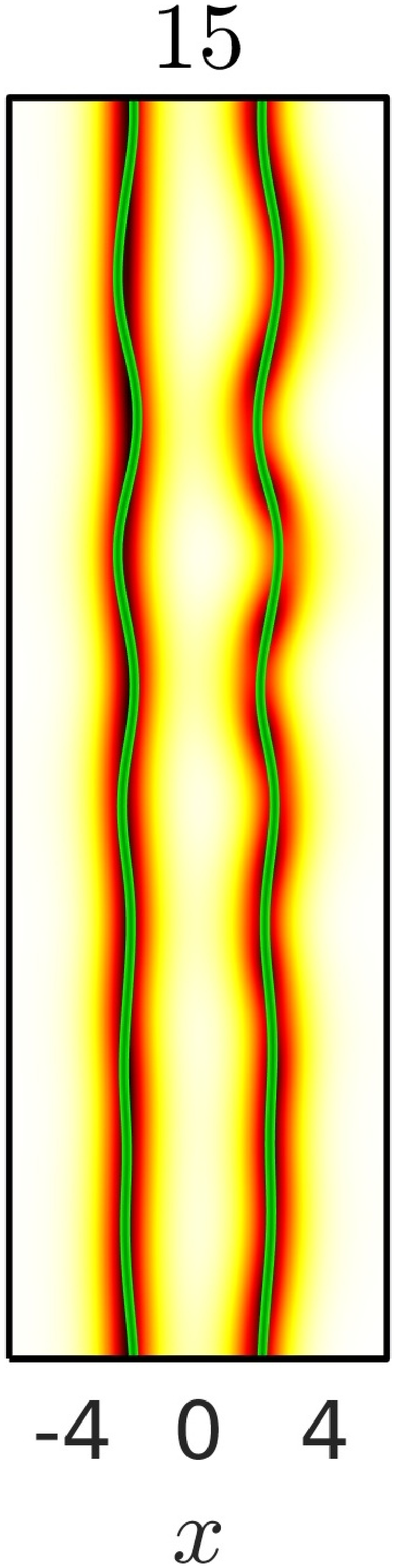}\myS
\includegraphics[height=\myH]{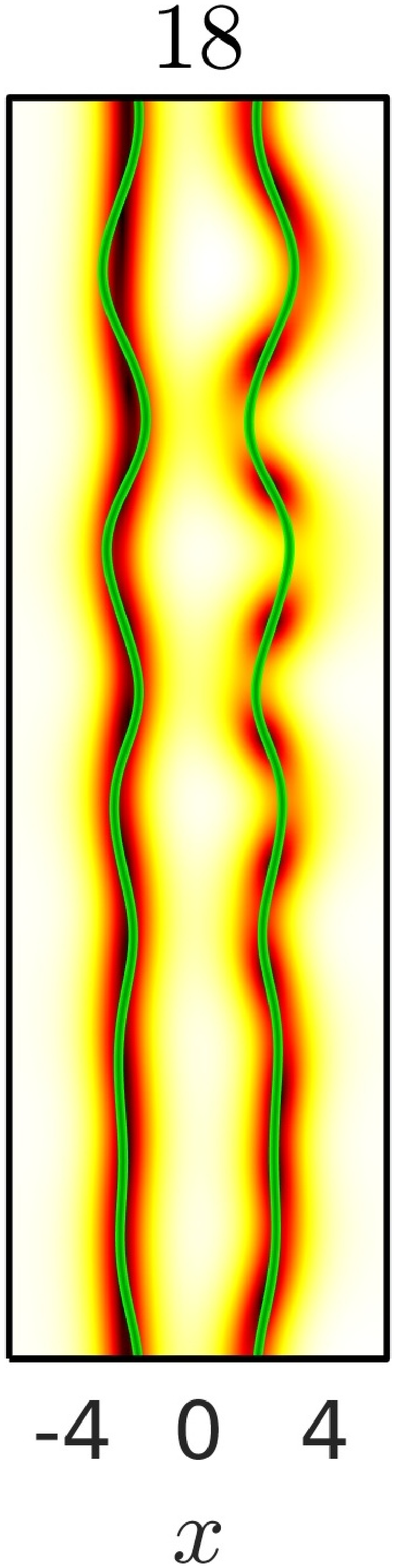}\myS
\includegraphics[height=\myH]{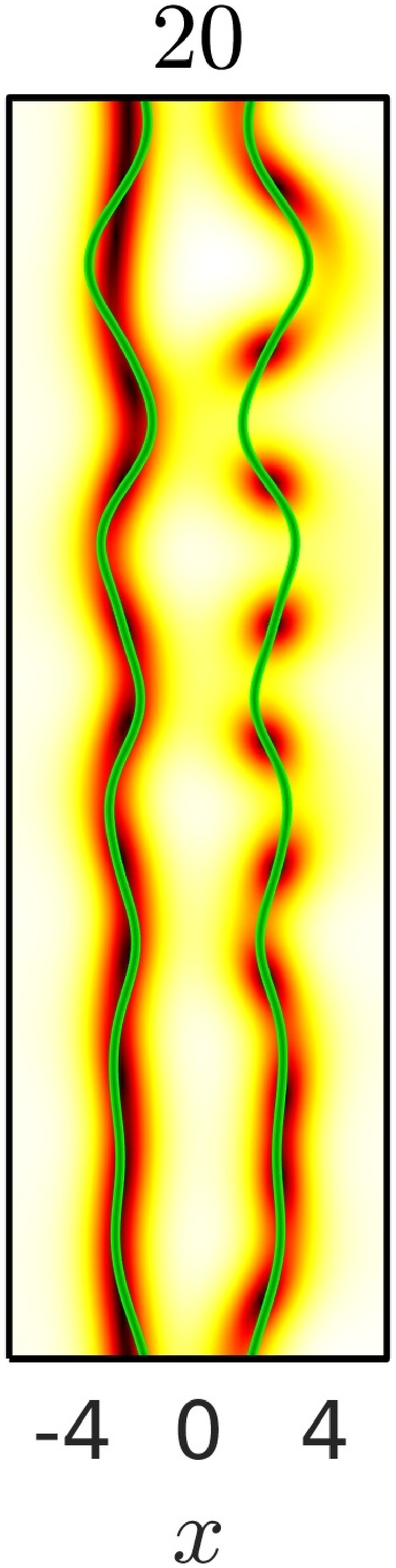}\myS
\includegraphics[height=\myH]{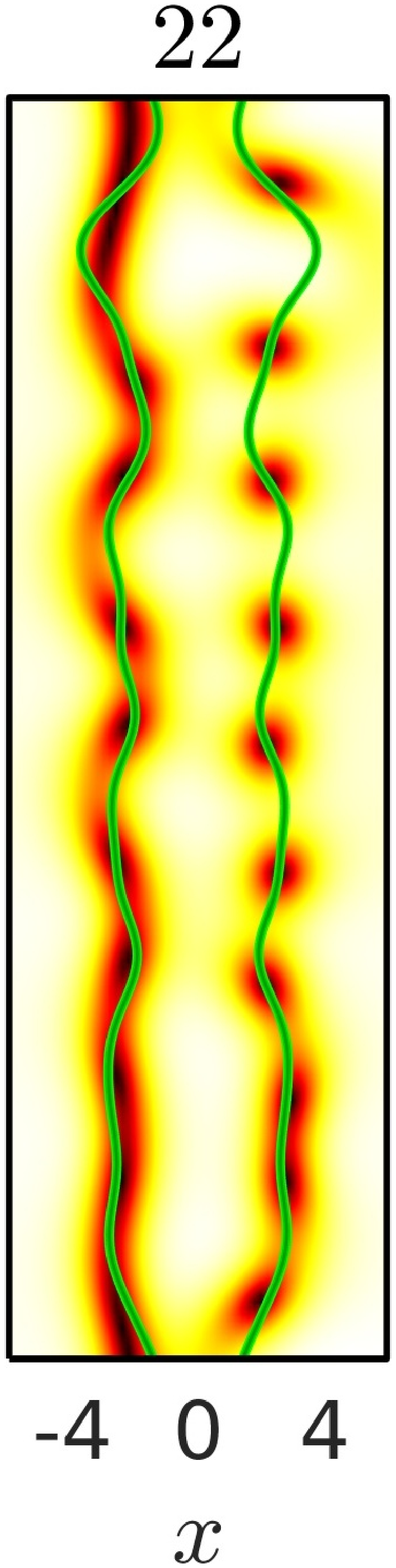}\myS
\includegraphics[height=\myH]{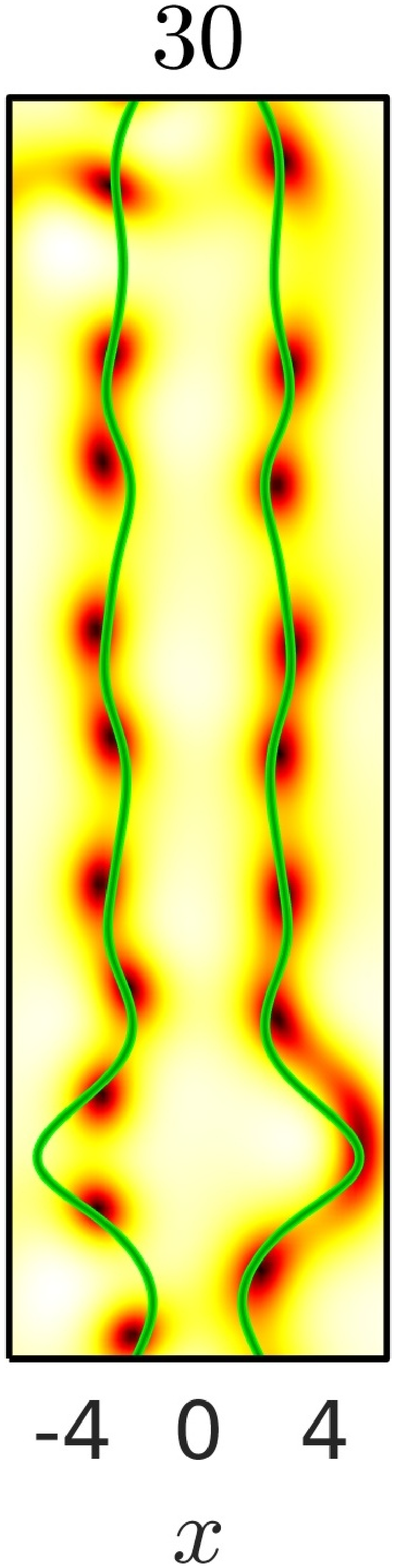}\myS
\includegraphics[height=\myH]{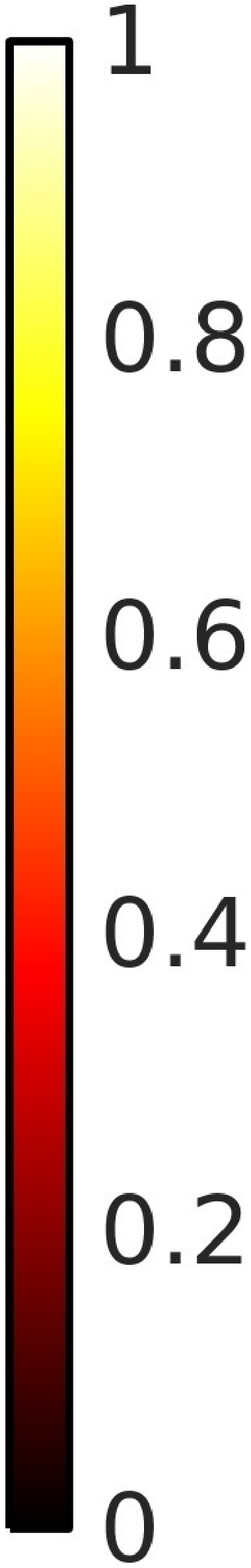}
\\ \vspace{0.2cm}
\includegraphics[height=\myH]{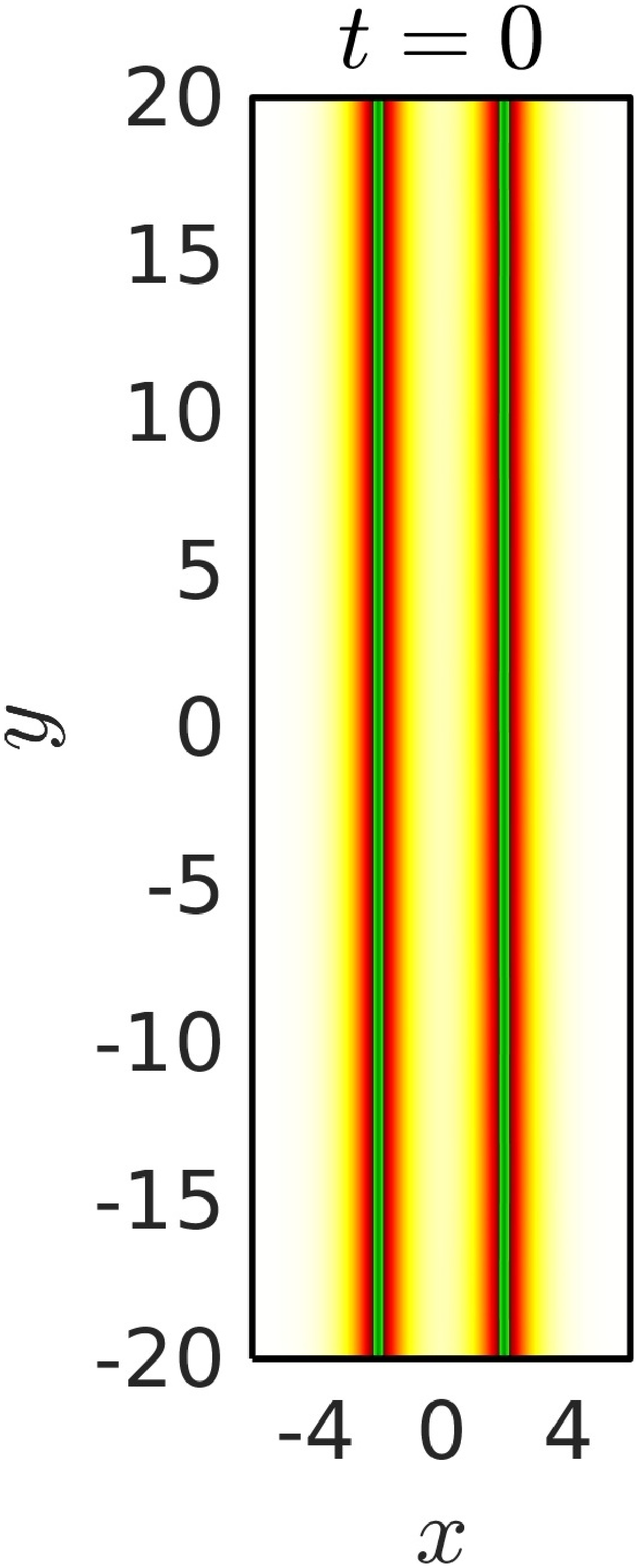}\myS
\includegraphics[height=\myH]{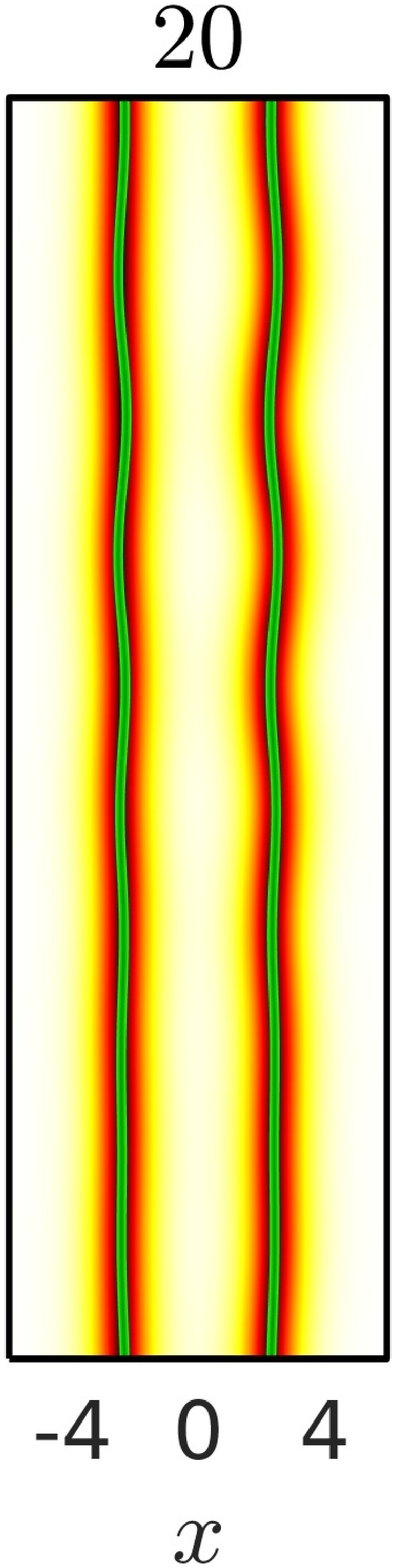}\myS
\includegraphics[height=\myH]{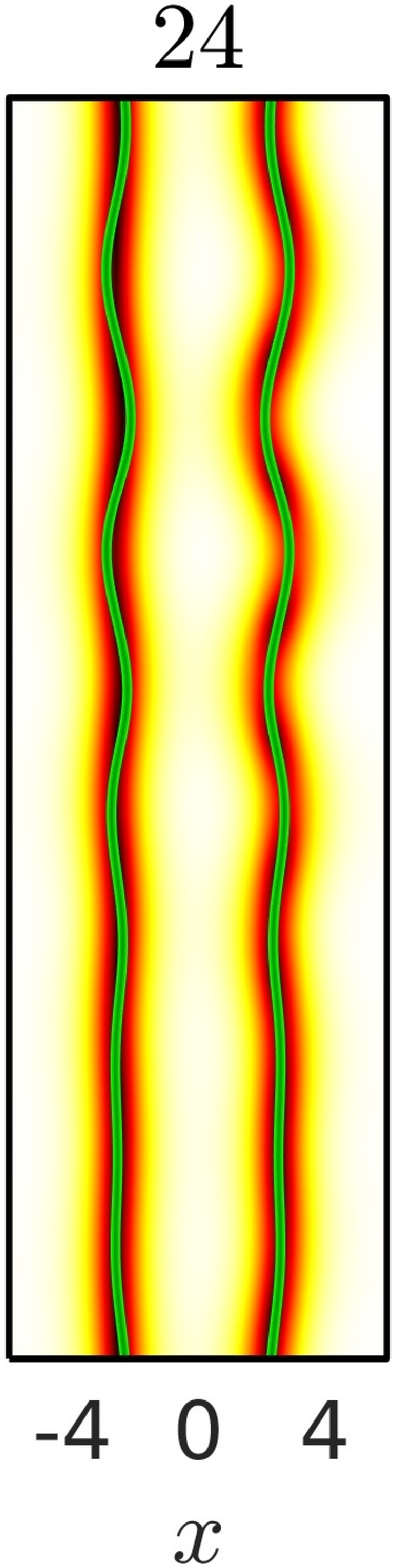}\myS
\includegraphics[height=\myH]{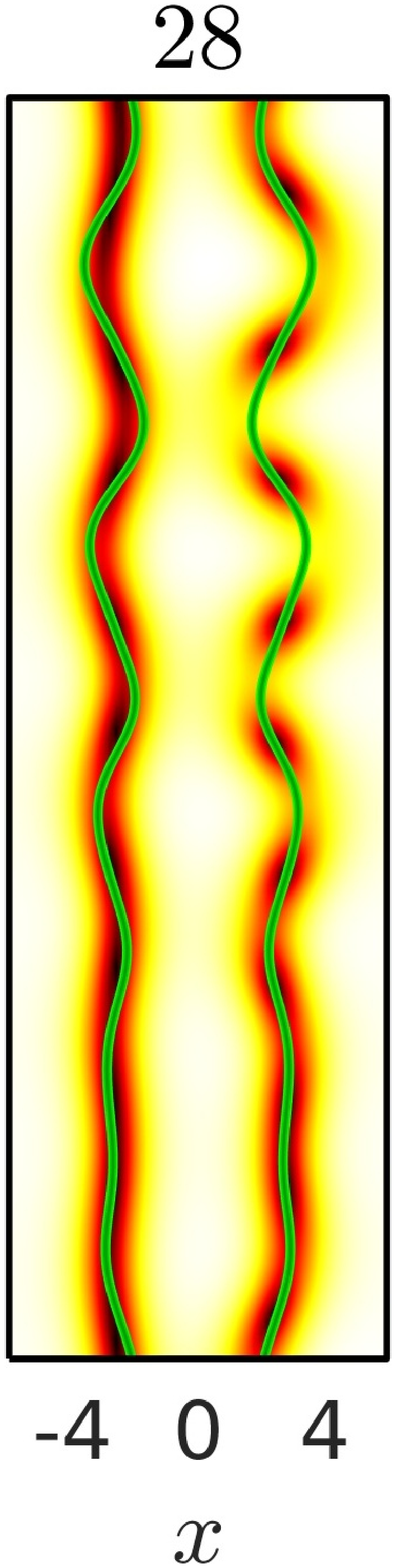}\myS
\includegraphics[height=\myH]{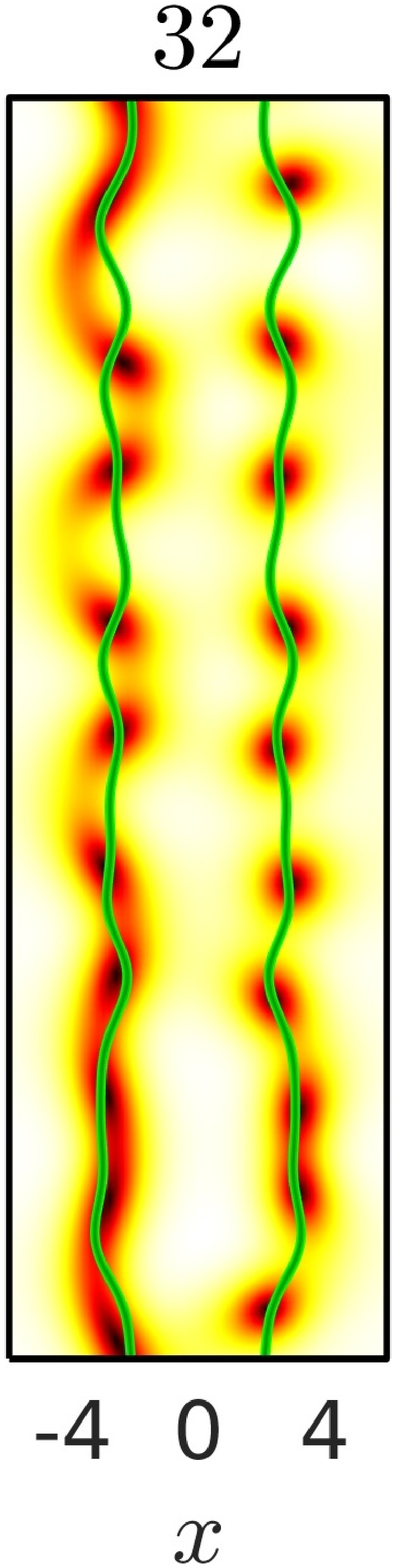}\myS
\includegraphics[height=\myH]{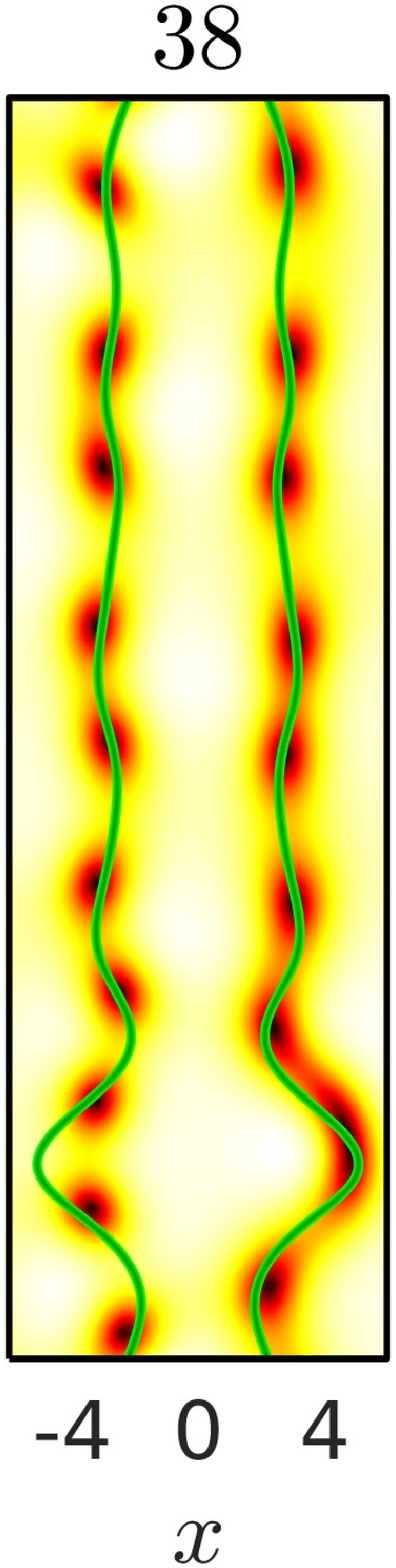}\myS
\includegraphics[height=\myH]{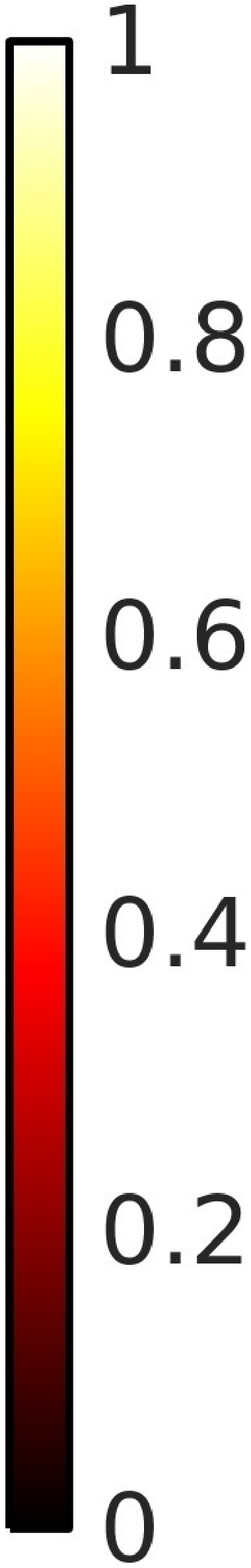}
\caption{(Color online) 
Snaking dynamics of two DSSs for $\mu=1$ perturbed by a 
non-symmetric perturbation where only the right DSS is perturbed.
Similar as in Fig.~\ref{fig:2snakes1-10} but for an {\em unperturbed}
left DSS given by $X_1(y,t=0)=-x_0$ and a right DSS perturbed by
$X_2(y,t=0)=x_0+\sum_{j=1}^{10}\varepsilon_j\,\sin[k\pi(y+\varphi_j)/L_y]$
with $x_0=2$, $\varphi_j=(j-1)L_y\pi/5$, and $k=j$
(i.e., same as in Fig.~\ref{fig:2snakes1-10}), where
the perturbation strengths for the right DSS are:
$\varepsilon_j=0.01$ (top series of panels) and
$\varepsilon_j=0.001$ (bottom series of panels).
%(middle series of panels). 
%and
%$\varepsilon_j=0.0001$ (bottom series of panels).
%
Note that although the dynamics for both cases is very similar, the
case with the weakest perturbation (see bottom panels) displays a 
perturbation growth at a slower time scale (contrast the difference 
in times between the two cases). Importantly also note the deviation
induced by the asymmetry between the (by construction
symmetric) VA and the full dynamics.
See Supplemental Material {\tt movie-2snakes-0-0-1-10-A}
and {\tt movie-2snakes-0-0-1-10-B} for animations 
depicting the corresponding dynamics.
}
\label{fig:2snakes0-0_1-10}
\end{center}
\end{figure}
%%%%%%%%%%%%%%%%%%%%%%%%%%%%%%%%%%%%%%%%%%%%%%%%%%%%%%%%%%%%%%%%%%%%%%%

We now proceed to test the improved VA prediction for the interaction 
of two DSSs. Similar to the one DSS case, let us probe both the long
wavelength scenario and multiple, mixed mode case.
Figure~\ref{fig:2snakes1} depicts the evolution for two stationary
(side-by-side) DSSs symmetrically perturbed by the longest
possible wavelength in the provided domain.
As the picture shows, the VA is able to track extremely well the DSS
dynamics up to the point where the DSSs touch, reconnect, and
split into patterns involving vortical structures.
It is important to note that we have chosen a configuration where
the interactions between the two stripes are quite not trivial. This can be
noticed by the fact that the top portions of the DSSs (for $y\approx8$)
initially get closer to each other (due to the individual growth rates of
the snaking instability for each DSS) and then {\em repel} each other 
($t>20$) once the DSS proximity is such that
the mutual dark soliton repulsion dominates the dynamics.
A more compelling case can be made by testing the VA prediction for the two 
DSSs by starting with an initial condition containing a {\em non-trivial}
combination of the first ten modes; see Fig.~\ref{fig:2snakes1-10}.
As for the one DSS case presented in Fig.~\ref{fig:snake1-10},
the two DSS VA reduction is also able to adequately track the two DSSs
even well after they break up into individual vortices; see in
particular the snapshots between $t=18$ and $t=26$.
It is relevant to mention that, as in single DSS case, two DSSs case
may also be approximated by the AI methodology~\cite{aipaper3}.
However, one needs to keep in mind that the AI approach is valid
for cases containing long wavelengths. Thus, for cases containing
shorter wavelengths (larger wavenumbers),
as it is particularly the case depicted in 
Fig.~\ref{fig:2snakes1-10}, the VA is a very good approximation
while the AI will fail in a similar manner akin to the results
presented in Fig.~\ref{fig:snake1-10} for the single DSS.

Finally, for completeness, we briefly study the case where the
perturbation on each of the two DSSs is not symmetric. In this case,
due to the chosen ansatz (containing a shared velocity term between
the two DSSs), the VA dynamics necessary leads, by construction, to 
a symmetric configuration. Therefore, in principle, one would not 
expect the VA to give a meaningful prediction for the two DSSs when
perturbed asymmetrically. Nonetheless, we tested that for perturbations
with small asymmetries the VA does indeed reasonably well at describing
the evolution of the two DSSs, despite its obvious shortcoming
in that the evolution of the full dynamics is no longer symmetric.
As an example, we depict in 
Fig.~\ref{fig:2snakes0-0_1-10} a couple of cases  where the left
DSS is left unperturbed $X_1(y,t=0)=-x_0$ while the right DSS is
perturbed as in the previous numerical examples. In particular, 
Fig~\ref{fig:2snakes0-0_1-10} depicts two case where the right soliton
position is perturbed by a linear combination of the first ten modes with (weak)
strengths of $0.01$ (top series of panels) and $0.001$ (bottom series 
of panels).
As the figure shows, after some transient, the original NLS (\ref{eq:NLS_2D})
dynamics evolves such that the left DSS develops undulations (exerted by the 
right DSS) that are in fact approximately symmetric with respect to the 
undulations of the right DSS. 
Therefore, it is not completely surprising that this behavior
is, to some extent, followed by the VA as
the results in Fig.~\ref{fig:2snakes0-0_1-10} show.

%%%%%%%%%%%%%%%%%%%%%%%%%%%%%%%%%%%%%%%%%%%%%%%%%%%%%%%%%%%%%%%%%%%%%%%%%%%%%
\section{Conclusions}
\label{sec:conclusions} 
%%%%%%%%%%%%%%%%%%%%%%%%%%%%%%%%%%%%%%%%%%%%%%%%%%%%%%%%%%%%%%%%%%%%%%%%%%%%%

In the present work we deployed the VA methodology in
order to provide an improved description of the transversely
unstable dynamics of single and multiple (two symmetrically perturbed)
DSSs in the defocusing two-dimensional
nonlinear Schr\"odinger equation.
%In particular, we applied the
%VA to a single DSS and to the dynamics of two, symmetrically perturbed, 
%interacting DSSs.
%
The method consists of applying the VA to suitable ans{\"a}tze that
consist of one or two DSSs whose transverse position and velocity
are functions of time and the transverse spatial variable. In this
manner, we are able to reduce the original (2+1)D NLS into two
coupled PDEs in (1+1)D for the dark soliton's position and velocity.
It is also important to highlight here that the standard VA
is qualitatively adequate, yet quantitatively misses the spectral
instability features of the single DSS. In view of that, we have
proposed an amended variant of the VA which captures the critical
wavenumbers $k_c$ of transverse restabilization (for different speeds) 
and consequently performs quantitatively better over the entire
range of wavenumbers. 
Our numerical results indicate that for short wavenumbers the VA
does a slightly better job at predicting the growth rates of
perturbations compared to the adiabatic invariant technique
developed in Refs.~\cite{aipaper,aipaper2,aipaper3}.
However, it is for perturbations containing larger wavenumbers 
(including a non-trivial mix for short and long wavenumbers)
that the advantage of the VA methodology is more apparent.
Specifically, the reduced equations of motion obtained through the VA
are able to closely track the dynamics for one and two interacting DSSs.
In fact, the reduced VA methodology is not only successful in tracking
the DSSs dynamics up to the point where the DSSs break up into vortices;
surprisingly, it continues to adequately trace the position
of the remnants of the former stripe even for times slightly beyond
its destabilization and breakup into vortical structures.
Our numerical results suggest that the reduced VA equations constitute a
viable methodology for theoretically reducing [to a (1+1)D setting]
and numerically closely tracking the dynamics of DSSs over a wide 
range of dynamical scenarios (i.e., containing a wide range of
perturbations from short to long wavelengths ones).

It is worth mentioning that the VA methodology seems not to be tractable
for a general two dark soliton ansatz. Therefore, in our presentation
we focus on a more specific ansatz where the two dark solitons share
velocities and are thus pushing to stay symmetric (each one being the
mirror image of the other one).
It would be interesting to generalize the results presented herein
by a suitable, and tractable, VA (or other) methodology
that would be able to successfully track two DSSs for arbitrary positions
and velocities. This would allow to not only treat the general two DSSs
case, but the general case of $N$ interacting stripes.
Of course, there are other directions that are relevant to pursue
as well, based on the program that has been recently developed at
the level of the adiabatic invariant methodology. Some of them concern ring dark
soliton structures, multi-component (e.g.~dark-bright) solitonic
patterns, as well as three-dimensional states such as planar or
spherical solitons. From a theoretical perspective, understanding
better how to justify an amendment like the one proposed herein
from first principles (that significantly improves the quantitative
tracking of the transversely unstable modes) is also an important
challenge for future studies.

\acknowledgments 

L.A.C.A.~gratefully acknowledges the leave of absence from IPN and the
financial support of the Fulbright-Garc\'ia Robles program during his
research stay at SDSU where part of this work was done.
R.C.G.~gratefully acknowledges the support of NSF-PHY-1603058.
P.G.K.~gratefully acknowledges the support of
NSF-PHY-1602994.
Constructive discussions with D.~Frantzeskakis, W.~Wang, B.~Anderson, 
G.~Theocharis, B.~Malomed and D.~Pelinovsky are gratefully acknowledged.

\end{document}